\documentclass{aa}  

\newcommand{\kms}{$\rm km\,s^{-1}$} 
\newcommand{\tco}{$\rm ^{13}CO$}
\newcommand{\cdo}{$\rm C^{18}O$}
\newcommand{\cdso}{$\rm C^{17}O$}
\newcommand{\tcdo}{$\rm ^{13}C^{18}O$}
\newcommand{\dtc}{$\rm ^{12}C/^{13}C$}
\newcommand{\dsdo}{$\rm ^{16}O/^{18}O$}
\newcommand{\dods}{$\rm ^{18}O/^{17}O$}

\usepackage{graphicx}
\usepackage{natbib}

\usepackage{txfonts}
\begin{document}

   \title{Spatially resolved carbon and oxygen isotopic ratios in NGC~253 using optically thin tracers.}
   \titlerunning{C and O isotopic ratios in NGC~253}
   
   \author{S. Mart\'in \inst{1,2}
          \and
          S. Muller\inst{3}
          \and
          C. Henkel\inst{4,5}
          \and
          D. S. Meier\inst{6,7}
          \and
          R. Aladro\inst{4}
          \and
          K. Sakamoto\inst{8}
          \and 
          P. P. van der Werf\inst{9}
          }

   \institute{European Southern Observatory, Alonso de C\'ordova, 3107, Vitacura, Santiago 763-0355, Chile 
              \email{smartin@eso.org}
         \and
             Joint ALMA Observatory, Alonso de C\'ordova, 3107, Vitacura, Santiago 763-0355, Chile
         \and 
             Department of Space, Earth and Environment, Chalmers University of Technology, Onsala Space Observatory, SE-43992 Onsala, Sweden
         \and 
             Max-Planck-Institut für Radioastronomie, Auf dem Hugel 69, 53121, Bonn, Germany
         \and
             Astron. Dept., King Abdulaziz University, P.O. Bon 80203, Jeddah 21589, Saudi Arabia 
        \and
             New Mexico Institute of Mining and Technology, 801 Leroy Place, Socorro, NM, 87801, USA
         \and 
             National Radio Astronomy Observatory, PO Box O, 1003 Lopezville Road, Socorro, New Mexico 87801, USA
         \and
	         Institute of Astronomy and Astrophysics, Academia Sinica, PO Box 23-141, 10617 Taipei, Taiwan
         \and 
         	Leiden Observatory, Leiden University, P.O. Box 9513, NL-2300 RA Leiden, the Netherlands
         }

   \date{}
 
  \abstract
{One of the most important aspects of modern astrophysics is related to our understanding of the origin of elements and chemical evolution in the large variety of astronomical sources.
Nuclear regions of galaxies undergo heavy processing of matter, and are therefore ideal targets to investigate matter cycles via determination of elemental and isotopic abundances.
}
{To trace chemical evolution in a prototypical starburst environment, we spatially resolve the carbon and oxygen isotope ratios across the central molecular zone (full size $\sim 600$~pc) in the nearby starburst galaxy NGC 253.
}
   {We imaged the emission of the optically thin isotopologues \tco, \cdo, \cdso, \tcdo~ at a spatial resolution $\sim50$~pc, comparable to the typical size of giant molecular associations. Optical depth effects and contamination of \tcdo~ by $\rm C_4H$ is discussed and accounted for to derive column densities.}
   {This is the first extragalactic detection of the double isotopologue \tcdo. 
   Derived isotopic ratios \dtc$\sim21\pm6$, \dsdo$\sim130\pm40$, and \dods$\sim4.5\pm0.8$ differ from the generally adopted values in the nuclei of galaxies.
}
   {The molecular clouds in the central region of NGC~253 show similar rare isotope enrichment
   to those within the central molecular zone of the Milky way. This enrichment is attributed to stellar nucleosynthesis. Measured isotopic ratios suggest an enhancement of $^{18}$O as compared to our Galactic center, which we attribute to an extra $^{18}$O injection from massive stars. Our observations show evidence for mixing of distinct gas components with different degrees of processing. We observe an extra molecular component of highly processed gas on top of the already proposed less processed gas being transported to the central region of NGC~253.
   Such multicomponent nature and optical depth effects may hinder the use of isotopic ratios based on spatially unresolved line to infer the star formation history and/or initial stellar mass function properties in the nuclei of galaxies.
}
   \keywords{ISM: molecules - ISM: abundances - Galaxies: abundances -  Galaxies: individual: NGC~253 - Galaxies: starburst - Submillimeter: ISM}

   \maketitle

\section{Introduction}
Interstellar isotope ratios carry essential information on the processes of nucleosynthesis
in the hot and dense interior of stars.
Measuring isotopic abundances such as \dtc~ and \dods~  in the nuclear regions of starburst galaxies can therefore reveal the fingerprint of their star formation history. 
Isotopic ratios can be also be related to the relative contribution of high-mass to intermediate-mass stellar processing \citep{Zhang2015}. Thus they can serve as tracers of the chemical evolution of the interstellar medium (ISM) due to stellar processing.
The recent work by \citet{Romano2017} attempts to model the available isotopic ratios observed in galaxies as a function of the star forming history and the stellar initial mass function (IMF).
Based on their comparison of observations and model results
they find evidence of a top-heavy initial mass function (IMF) being responsible for the observed ratios in starburst galaxies.

However, measuring isotopic ratios in astronomical sources is not always straightforward. At optical wavelengths, such measurements are difficult or impossible due to the blending of atomic isotope lines and isotope specific obscuration due to dust grains. Isotopic measurements of atomic carbon in the far-infrared and carbon monoxide in the near-infrared have been obtained \citep[i.e.][and references therein]{Goto2003}. On the other hand, observations of molecular isotopologues at radio to submillimeter wavelengths have been proven most successful within the Galaxy, towards the local extragalactic ISM \citep[see][and references therein]{Mart'in2010a,Henkel2014,Romano2017}, and all the way to high redshift molecular absorbers \citep[i.e.,][]{Muller2011,Wallstroem2016}.

In this paper we focus on the isotopic ratio between $^{12}$C and $^{16}$O, as primary products of stellar nucleosynthesis, and the rarer $^{13}$C, $^{18}$O, and $^{17}$O, as secondary nuclear products from primary seeds \citep{Meyer1994,Wilson1992,Wilson1994}. 
We note that, primary production of  $^{13}$C  and $^{18}$O is predicted for fast rotating low metallicity massive stars and intermediate-mass stars climbing the asymptotic giant branch \citep{Chiappini2008,Karakas2014,Limongi2018}, but the former results in a relatively small enrichment while the latter will not be relevant for the overall galactic scales involved in this work.
In this context, when referring to enrichment we mean rare isotopologues enrichment, and therefore lower \dtc,\dsdo, and \dods~ isotopic ratios.
As compiled in \citet{Wilson1994} and \citet{Henkel1994a}, the generally adopted values in the nuclei of low luminosity local starbursts are $\rm^{12}C/^{13}C\sim40$, $\rm^{16}O/^{18}O\sim200$, and $\rm^{18}O/^{17}O\sim8$ \citep{Sage1991,Henkel1993,Henkel1993a}.
These ratios are derived from C- and O-bearing molecular species commonly observed in the ISM, such as CO, CN, CS, HCN, HNC, and HCO$^+$, which, despite not suffering from the line blending between the main and rare isotopologues, may actually be biased by optical depth effects. These limitations result in most of the cited ratios to be lower limits. To circumvent this problem, it is sometimes possible to turn to rare isotopologues involving another element, for example studying the double ratio $^{12}$C$^{34}$S/$^{13}$C$^{32}$S, but this adds potential problems due to poor understanding of the extra isotopic ratio \citep[in this example, $^{32}$S/$^{34}$S,][]{Henkel2014,Meier2015}. 

Recently, \citet{Henkel2014} used high quality spectra of CN towards the starburst galaxy NGC~253 to revisit its $^{12}$C/$^{13}$C ratio. Thanks to the hyperfine splitting of CN, it is possible to have a good estimate of the optical depth of the main isotopologue. The resulting  $\rm^{12}C/^{13}C\sim40\pm10$ confirmed some of the previous estimates. The accurate measurement of  the $\rm^{12}C/^{13}C$ ratio also becomes important because this ratio is often used to derive other atomic isotopic ratios based on observations of the optically thinner $^{13}$C bearing isotopologues \citep[i.e.,][]{Martin2005}.

However, there are several non local thermodynamic equilibrium (non-LTE) effects that can affect the observed molecular line ratios based on optically thick species \citep{Meier2001,Meier2015}. As pointed out by \citet{Wilson1994}, in order to minimize effects such as optical thickness of the emission, excitation differences due to resonant photon trapping, and/or selective photodissociation, one would rather choose the observation of optically thin isotopologues such as \cdo~ and \tcdo. Despite the weakness of the emission of these isotopologues, observations have been successfully performed across the Galaxy \citep{Langer1990,Ikeda2002}, but are very challenging in extragalactic sources.
High sensitivity observations with the IRAM~30m telescope were carried out by \citet{Mart'in2010a} in an attempt to detect \tcdo~ towards the nearby starburst galaxies NGC~253 and M~82. The resulting tentative detection towards NGC~253 yielded a lower limit of $\rm^{12}C/^{13}C\gtrsim60$, comparatively higher than previous estimates.

Another important effect to take into account is that of isotopic fractionation \citep{Watson1977,Langer1984,Wilson1994,Roellig2013,SzHucs2014,Roueff2015} 
where the unbalance of exothermic isotope exchange chemical reactions may favor the enhancement of the rarer isotopologues. The most relevant isotope barrierless fractionation reactions for the carbon monoxide species studied in this paper are
\begin{equation*}
\begin{split}
^{13}{\rm C}^+ + {\rm CO} \rightarrow {\rm C}^+ + \ ^{13}{\rm CO} + \Delta E_1 \\
^{13}{\rm C}^+ + {\rm C}^{18}{\rm O} \rightarrow {\rm C}^+ + \ ^{13}{\rm C}^{18}{\rm O} + \Delta E_2\\
^{18}{\rm O^+} + {\rm CO} \rightarrow {\rm O}^+ + \ \rm C^{18}{O} + \Delta E_3 \\
\end{split}
\end{equation*}
where the heats of the reaction are $\Delta E_1 = 35$~K, $\Delta E_2 = 36$~K, and $\Delta E_3 = 38$~K, respectively \citep{Langer1984,Loison2019a}. The equilibrium constant, and therefore the relation between the measured and observed isotopologue ratios is proportional to $\exp(\Delta E/T_{kin})$, where $T_{kin}$ is the kinetic temperature of the gas \citep{Wilson1994}. Therefore, these reactions are relevant in the cold ISM.

Based on the observed gradients in the Galaxy by \citet{Milam2005}, \citet{Romano2017} concluded that carbon fractionation effects are negligible for the bulk of the gas in galaxies. However, \citet{Jimenez-Donaire2017} invoke fractionation as contributor to the \tco/\cdo~ gradient with galactocentric radius in a sample of nearby galaxies.

Given that the central molecular zone of NGC~253 has a relatively high kinetic temperature \citep{Ott2005,Mangum2013,Mangum2018,Krips2016}, fractionation effects should be negligible in the gas phase chemistry. Since we are not able to provide sensible constraints on fractionation issues in the framework of this paper, we will therefore simply assume that our measured isotopologue ratios reflect true isotopic ratios. 

The bright molecular emitting central molecular zone (CMZ) of the nearby starburst galaxy NGC~253 \citep{Mart'in2006,Aladro2015} is ideally suited for high resolution studies of the isotopic ratio distribution in an extragalactic environment. At a distance of $\sim3.5$~Mpc \citep{Rekola2005,Mouhcine2005}, it can be easily resolved by interferometric observations at giant molecular cloud scales of a few tens of pc.
Within the CMZ of the Milky way, inhomogeneities in the isotopic ratio distribution have been found and attributed to accretion of material from the outer disk \citep{Riquelme2010}. 

In this paper, we aim at shedding light on discrepancies between the ratios derived from low resolution single dish observations, and to study the spatial distribution of the carbon and oxygen isotopic ratios in an extragalactic starburst region. We present high angular resolution and high sensitivity observations of four rare carbon monoxide isotopologues, \tco, \cdo, \cdso, \tcdo, reporting the first firm detection of the double isotopologue \tcdo~ in an extragalactic environment.
 
\section{Observations}
\label{Sec.Observations}
The observations were carried out with the Atacama Large Millimeter and Submillimeter Array (ALMA) under the Cycle 4 project 2016.1.00292.S (P.I. S. Mart\'in). The observations consisted of two different receiver tunings.
The first tuning setup had spectral windows centered at 91.0
, 92.8, 102.9, and 104.7~GHz and aimed at covering the $J=1-0$ transition of $\rm ^{13}C^{18}O$ at 104.711~GHz.
The second tuning setup had spectral windows centered at 97.9, 99.6, 110.0,
and 111.9 GHz and aimed to cover the $J=1-0$ transitions of $\rm ^{13}CO$ (110.201~GHz), $\rm C^{18}O$ (109.782~GHz), and $\rm C^{17}O$ (112.359~GHz).
In both cases the spectral windows were configured to cover a bandwidth of 1.875~GHz each with a channel spacing of 0.488~MHz (corresponding to $\sim1.3-1.4$~\kms\ velocity resolution) after Hanning smoothing.

\begin{figure*}
\centering
\includegraphics[width=\hsize]{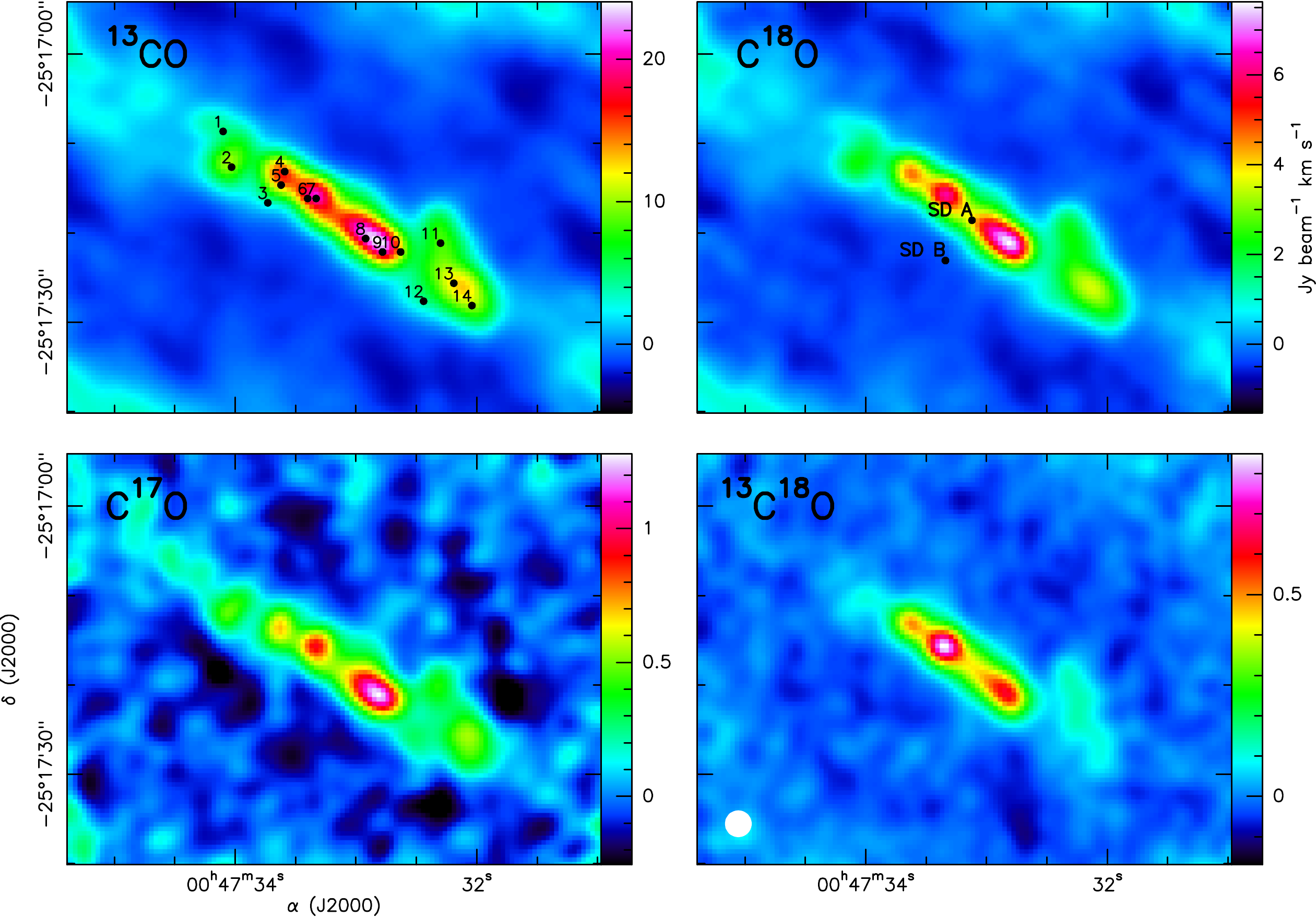}
   \caption{Integrated intensity images of the $J=1-0$ transitions of all the CO isotopologues targeted in this work. Line emission is integrated in the velocity range of  $50-450$~\kms. The common beam size of $3''$ ($\sim50$~pc) is shown as a white circle in the bottom left corner of the \tcdo~ map. Note that the emission of \tcdo~ is contaminated at the higher velocities in some positions as discussed in Sect.~\ref{Sect.LineContamination} and shown in Fig.~\ref{fig.linecontamination}. This contamination results in the brighter \tcdo~ region at position 7 differing from the brightest spot from the other isotopologues approximately at position 9. Marked positions numbered 1 to 14 indicate those selected for spectra extraction and analysis (Sect.~\ref{sec.spectra}). Nominal positions of the single dish spectra from the literature used for comparison are labeled SD A and B (Sect.~\ref{Sec.SpatialFiltering}).}
   \label{fig.maps}
\end{figure*}

Observations consisted of a single pointing at the nominal phase center of $\alpha_{J2000}=00^{\rm h}47^{\rm m}33\fs182$, $\delta_{J2000}=-25^\circ17'17\farcs148$.
The field of view of $\sim60''$ at these frequencies is wide enough to cover the whole central molecular zone in NGC~253 (see Fig.~\ref{fig.maps}), roughly $\sim40''\times10''$ ($\sim 650$~pc~$\times150$~pc) in size.
The observation targeting the faint $\rm ^{13}C^{18}O$ transition were carried out between January 1st and 3rd, 2017, with a total on-source time of 170 minutes and unprojected baselines ranging $15-460$~m  ($5.2-160~\rm k\lambda$) which results in an estimated maximum recoverable scale of $\sim15''$.
The precipitable water vapor (PWV) conditions were between 4--6~mm during these observations.
The second tuning, targeting mainly the brighter $\rm C^{18}O$ transition, was observed on December 23th, 2016 with only 12 minutes of on-source integration and unprojected baselines ranging $15-492$~m ($5.5-180~\rm k\lambda$) resulting in maximum recoverable scales of $13''$. The PWV was $\sim 2.5$~mm. In all cases, the number of 12~m-antennas in the array was between 43 and 46.

Observations of the quasars J0006$-$0623 and J0038$-$2459 were performed for bandpass and complex gain calibration, respectively. Absolute flux calibration used observations of Neptune and the quasar J0038$-$2459.
Based on the fluxes derived for the bandpass and gain calibrators between the different dates, we conservatively estimate an uncertainty $<10\%$ between the two frequency setups.

Calibration and imaging was performed in CASA version 4.7.2 \citep{McMullin2007}. The continuum emission was subtracted from the visibilities (with the CASA task uvcontsub) using line free channels.

With similar ranges of baseline coverage, the resulting synthesized beams differ only by $\sim5\%$ between the two setups. Imaging after cleaning was performed with a common circular synthesized beam of $3''$ (FWHM) for the sake of accurate comparison of flux densities between transitions. This angular resolution is equivalent to $\sim 50$~pc at the distance of NGC~253.

Final datacubes of the individual transitions were smoothed to a common 10~\kms\ velocity resolution. The achieved rms sensitivities in the velocity smoothed cubes were $\sim1.5$ and $\sim5$~mJy~beam$^{-1}$, equivalent to $\sim19$ and $\sim62$~mK, for the first and second tuning, respectively.

\section{Results}

\subsection{Integrated flux density maps}
\label{Sec.Inegratedmaps}

All targeted CO isotopologues were detected at high signal-to-noise ratio 
over the whole central molecular zone in NGC~253. In Fig.~\ref{fig.maps} we show the integrated flux density maps of the $J=1-0$ transition of all four CO isotopologues. The map of C$^{17}$O in Fig.~\ref{fig.maps} at $3''$ is similar to the one presented by \citet{Meier2015}, but the 
higher spatial resolution allows for resolving the individual GMC complexes in the very central region. 

Apart from the difference in line intensities (and therefore also signal-to-noise ratios), all four isotopologues show a consistent distribution, except for \tcdo~ which differs on the position of the brighter emission. This is not due to the intrinsic distribution of \tcdo~ but to the contamination by other species (Sect.~\ref{Sect.LineContamination}). We tried to generate the \tcdo~ integrated map with a mask per channel based on the emission extent observed in \tco. Unfortunately, the closeness of the emission of the molecular contaminant at just $\sim 23$\kms~(Sect.~\ref{Sect.LineContamination}) from the \tcdo~transition made the de-blending impossible in the integrated map. Thus, we preferred to show and use the raw integrated map in the selected velocity range with this limitation in mind when interpreting the map ratios (Sect.~\ref{sec.ratiomaps}).

Carbon and oxygen isotopic ratios can be estimated using the ratio of the carbon monoxide isotopologues as a proxy (Sect.~\ref{sec.ratiomaps}).
However, due to the mentioned contamination of the \tcdo~emission, a more accurate measurement can be derived from precise modeling and de-blending of its emission from spectra extracted on selected positions (Sect.~\ref{sec.spectra}).

\subsubsection{Spatial filtering}
\label{Sec.SpatialFiltering}

\begin{figure}
\centering
\includegraphics[width=\hsize]{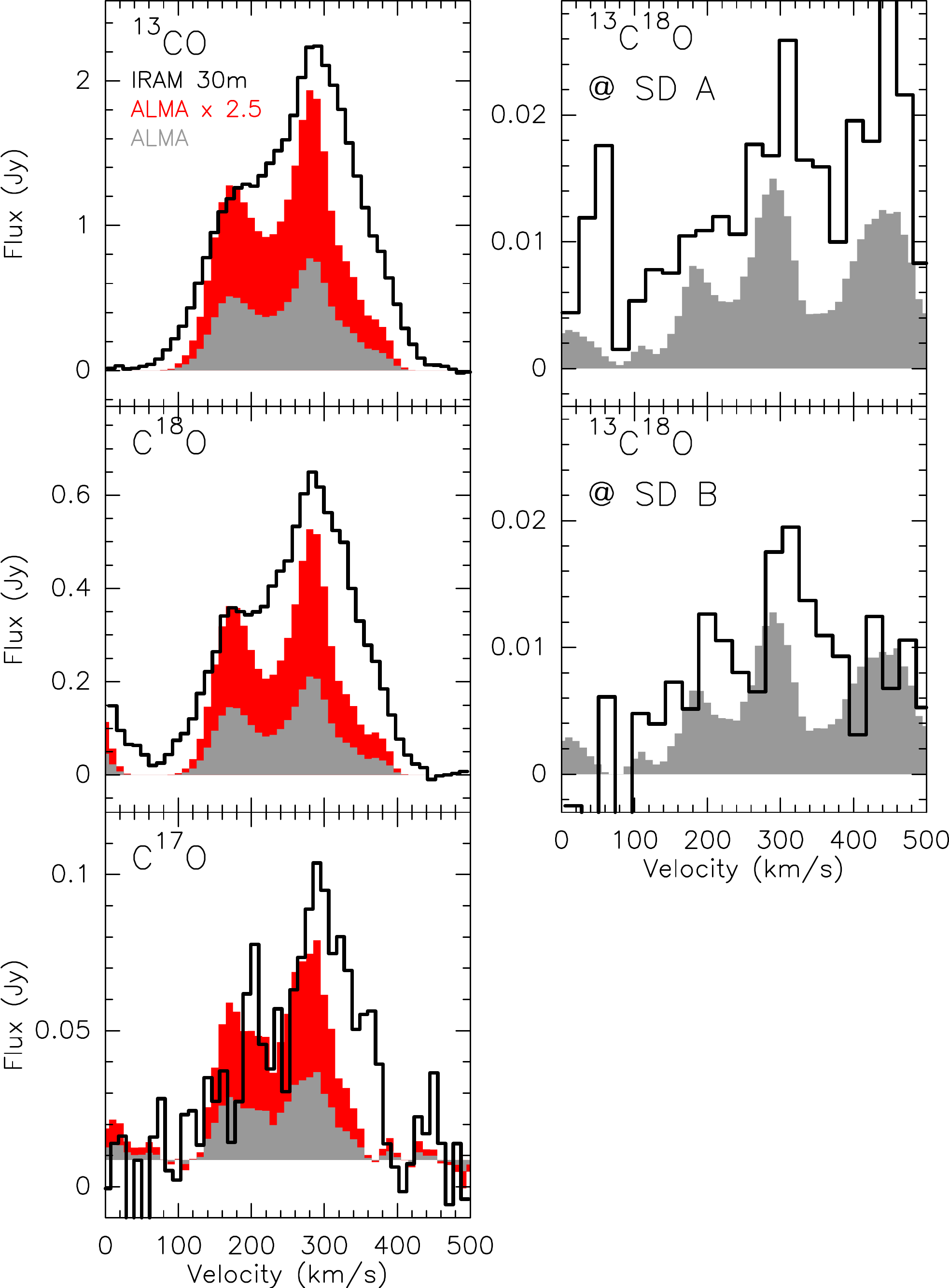}
   \caption{Spectra measured with the IRAM~30 telescope \citep{Aladro2015} and the ALMA observations presented in this work. ALMA spectra were extracted at the same positions as the single-dish observations from the data cubes smoothed to $23''$ resolution. For \tco, \cdo, and \cdso, the ALMA data are shown both with their actual flux (grey histogram) and multiplied by 2.5 (red histogram) for comparison with the single dish line profiles (black line). For \tcdo~, we show the profiles in both SD~A and SD~B positions in Fig.~\ref{fig.maps}, for comparison with the single dish data from \citet{Aladro2015} and \citet{Mart'in2010a}, respectively. See text in Sect.~\ref{Sec.SpatialFiltering} for details.}
\label{fig.missingflux}
\end{figure}

Despite using one of the most compact configurations of the ALMA 12m array for the sake of achieving a very low brightness temperature sensitivity (Sect.~\ref{Sec.Observations}), we still expect that a fraction of the flux is filtered out due to our lack of zero spacing observations.

To evaluate the missing flux due to spatial filtering, we reconstructed our ALMA primary beam corrected cubes with a $\sim23''$ beam, equivalent to that of the IRAM~30m telescope at the frequency of the CO isotopologues $J=1-0$ transitions.
We extracted the smoothed spectra at the same positions as those from the single dish observations.
These positions (indicated in Fig.~\ref{fig.maps}, upper right panel) are SD~A, at $\alpha_{J2000}=00^h47^m33\fs12$, $\delta_{J2000}= -25^\circ17'18\farcs6$, 
to compare with the spectra from \citet{Aladro2015} of the four observed isotopologues, and SD~B,
$\alpha_{J2000}=00^h47^m33\fs34$, $\delta_{J2000}= -25^\circ17'23\farcs1$, 
to compare with the \tcdo~ profile reported by \citet{Mart'in2010a}.
Extracted spectra on the smoothed cubes are shown in Fig.~\ref{fig.missingflux} as grey histograms.

Based on our smoothed datacubes we could verify that the published \tcdo~ integrated flux by \citet{Mart'in2010a} (position SD~B) would have been $25\%$ brighter if measured towards the position from \citet{Aladro2015} (position SD~A), which is closer to the galaxy center. This fact may have an impact in the \dtc~ ratio reported  by \citet{Mart'in2010a} as discussed in Sect.~\ref{Sec.IsotopicRatios}. Moreover, the limited pointing accuracy of the single dish observations may also introduce some uncertainties when comparing different datasets.

The IRAM~30m single dish spectra, shown in Fig.~\ref{fig.missingflux} in black lines, have been converted to flux density units with a factor of $S/T_{MB}=5$~Jy/K \footnote{http://www.iram.es/IRAMES/mainWiki/Iram30mEfficiencies}. Additionally, for \tco, \cdo, and \cdso, we plot the ALMA spectra (grey histograms) as extracted from the corresponding position in our $23''$ smoothed cubes, and also multiplied by a factor of 2.5 (shown in red) for the sake of line shape comparison with the single dish data.

We find that, integrated over the velocity range 40-450~\kms, 
only $\sim28\%$, $26\%$, and $29\%$ 
of the \tco, \cdo, and \cdso~ single dish flux is recovered by our ALMA data, respectively. These values are very similar and hint at similar spatial distributions as already pointed out in Sect.~\ref{Sec.Inegratedmaps} based on the maps in Fig.~\ref{fig.maps}. However, this is around a factor of 2 lower than the flux recovery of $60\%$ reported for \cdso~ by \citet{Meier2015}. 
Such difference is due to the different single dish integrated flux reported by \citet{Henkel2014} ($1.68\pm0.15$~K~\kms) that \citet{Meier2015} used as a reference and the factor of 2 larger value from \citet{Aladro2015} ($3.15\pm0.14$~K~\kms) we use in this paper. The different single dish fluxes are due to different observed pointing positions.
This is reflected by the narrower line width derived by \citet{Henkel2014}, mainly targeting the southwestern part of the central ridge, as compared to that by \citet{Aladro2015}. When using the same single dish reference, both \citet{Meier2015} and this work recover the same fluxes.
We consider our missing flux estimates more robust since they are derived from smoothed data at the same resolution and position as the single dish data.
\citet{Meier2015} reported scales sampled up to $18''$ (90th percentile), similar to our larger recoverable scales (Sect.~\ref{Sec.Observations}), which does point out to significant amounts of gas at larger scales.

For \tcdo, the recovered flux measured is $\sim60\%$ and $\sim45\%$ for the \citet{Mart'in2010a} and \citet{Aladro2015} positions, respectively.
If we consider velocities only up to 350~\kms~ to reduce the effect of line contamination at the higher  velocities (Sect.~\ref{Sect.LineContamination}), the recovered flux yields similar values of $\sim55\%$ and $\sim45\%$, respectively. Although, the longer integration at this frequency setup might result in a better inner-UV coverage, the difference in the recovered flux compared to the other isotopologues can be attributed to: significantly varying spatial distribution of \tcdo~, but, in principle, we would not expect it to be so different from the other rare isotopologues; a higher degree of compactness of the emission of the molecular contaminant (Sect.~\ref{Sect.LineContamination}) which contributes to the measured integrated emission; and more importantly, to the lower signal to noise of this spectral profile affected by both a larger absolute flux uncertainty due to the noise and the extra uncertainty due to the baseline subtraction performed on the single dish spectra (See Fig.~\ref{fig.missingflux}).


These results 
point to the uncertainties inherent to single dish observations. In that sense, as pointed out by \citet{Meier2015}, the interferometric observations may provide us with a more homogeneous look at the isotopic ratios since we are filtering out approximately the same extended emission in all transitions and therefore we are obtaining information from the same spatial scales. This is also evidenced by the similar line widths from all of our observed line profiles (e.g. in Fig.~\ref{fig.missingflux}). As mentioned above and followed up in Sect.~\ref{Sec.IsotopicRatios}, this issue poses some questions regarding the accuracy and interpretation of results obtained from low resolution and non-simultaneous heterogeneous observations.

Fig.~\ref{fig.missingflux} also shows that should \tcdo~ be affected by a similar flux filtering than the other isotopologues, assuming a correct baseline subtraction, it should have potentially been detected at higher signal-to-noise ratio in the IRAM~30m single dish data by both \citet{Mart'in2010a} and \citet{Aladro2015}.

\subsubsection{Line ratio maps}
\label{sec.ratiomaps}
\begin{figure}
\centering
\includegraphics[width=\hsize]{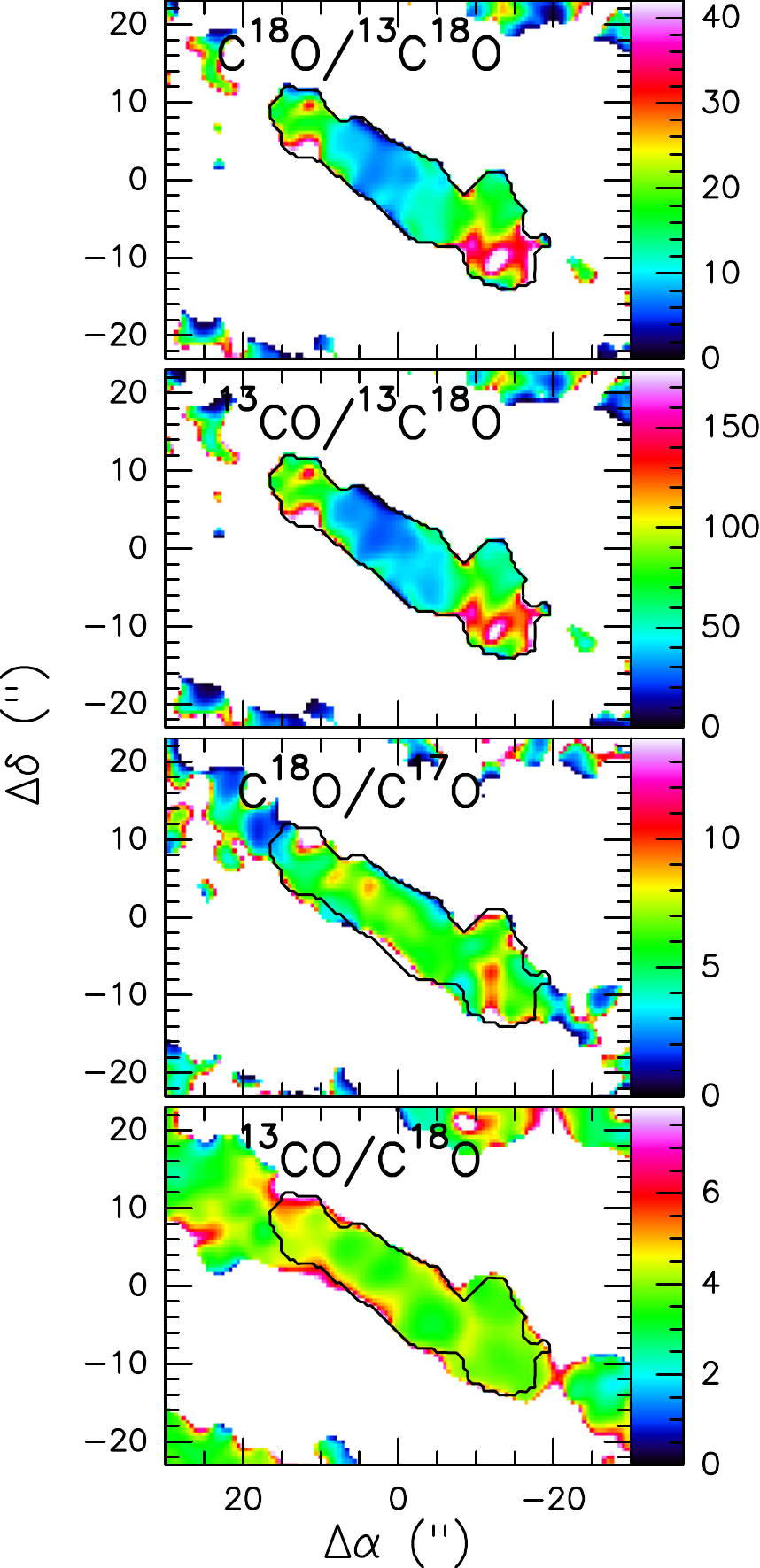}
   \caption{Integrated flux density ratio maps derived from the maps shown in Fig.~\ref{fig.maps}. Only the ratios discussed in the paper are shown. The black contour is the central $1~\sigma$ level of \cdo, used to derive ratio averages within the maps. See Sect.~\ref{sec.ratiomaps} for details. }
   \label{fig.mapsratios}
\end{figure}

In Fig.~\ref{fig.mapsratios} we show the integrated flux density ratio maps for the relevant pairs of transitions. These ratios have been calculated for pixels where the signal in the integrated maps in the $50-450$~\kms~range is $>1\sigma$, which is $\rm >30~mJy~beam^{-1}~km~s^{-1}$ in \tco, \cdo, and \cdso, and $\rm >9.5~mJy~beam^{-1}~km~s^{-1}$ for \tcdo.
The uncertainty in these ratio maps is described in Appendix~\ref{Sec.ErrorMapRatios}.
The first central \cdo\ contour fulfilling that criterion is displayed in both Fig.~\ref{fig.mapsratios} and Fig.~\ref{fig.errormapsratios}, and is considered to be the region of relevance for these ratios. Outside this area, although there is emission above the thresholds, we cannot ensure these are not residuals from the cleaning process.

The average and standard deviation within this region with different weighting schemes are shown in Table~\ref{tab.isotopiratios} for each line ratio. The color scale in Fig.~\ref{fig.mapsratios} was set from 0 to the unweighted average plus twice the standard deviation. 
On such color scale, the \cdo/\tcdo~ ratios appear to be very similar to those of \tco/\tcdo~ (upper two panels of Fig.~\ref{fig.mapsratios}), indicating a rather constant \tco/\cdo~ line intensity ratio as can also be seen in the lowest panel of Fig.~\ref{fig.mapsratios}. The \cdo/\tcdo~ and \tco/\tcdo~ ratios appear to show a gradient from the inner region towards the outskirts of the CMZ where some clumps show very high ratios. We note that the decrease of the ratio towards the center (bluish regions in the upper two panels of Fig.~\ref{fig.mapsratios}), is caused by the increased line contamination in \tcdo\ as seen in Fig.~\ref{fig.maps}. Thus, \cdo/\tcdo~ and \tco/\tcdo~ ratio maps as well as their derived averages in Table~\ref{tab.isotopiratios} need to be considered with caution in this region as discussed in Sect.~\ref{sec.mapvsspectra}.
\cdo/\cdso~ and \tco/\cdo~ appear somewhat smoother, and indeed their relative dispersion in Table~\ref{tab.isotopiratios} are lower than in the other ratios.

\begin{table*}
\caption{Averaged derived isotopic ratios in NGC~253 \label{tab.isotopiratios}}
\centering
\begin{tabular}{c c | c c c | c c c}
\hline
\hline
Isotopic        & Molecular   &\multicolumn{3}{c}{Maps \tablefootmark{a}}      & \multicolumn{3}{c}{Spectra \tablefootmark{b}}\\
ratio           &   ratio     & unweighted & $\sigma-$weighted & snr-weighted                     & unweighted & $\sigma-$weighted & $\tau-$weighted\\
\hline
$\rm^{12}C/^{13}C$   &  \cdo$/$\tcdo    & $18\pm12$ & $13\pm8$ & $11\pm5$      & $22\pm7$ & $21\pm6$ & $23\pm5$ \\
$\rm^{16}O/^{18}O$   &  \tco$/$\tcdo    & $75\pm52$ & $48\pm30$ & $40\pm20$    & $140\pm50$ & $130\pm40$ & $140\pm40$ \\ 
$\rm^{18}O/^{17}O$   &  \cdo$/$\cdso    & $6.5\pm3.7$ & $6.2\pm1.6$ & $5.7\pm1.6$     & $4.6\pm1.1$  &  $4.5\pm0.8$ & $4.7\pm0.8$ \\
                     &  \tco$/$\cdo     & $4.5\pm1.7$ & $3.8\pm0.6$ & $3.8\pm0.5$     & $6.2\pm1.0$  &  $6.1\pm0.9$ & $6.1\pm0.9$ \\
\hline
\end{tabular}
\tablefoot{
\tablefoottext{a}{Averages and standard deviations derived from flux density ratio maps calculated within the region of interest in Fig.~\ref{fig.mapsratios}. See Sect.~\ref{sec.ratiomaps} for details. Parameters have been calculated unweighted, as well as weighted by the $\sigma$ and signal-to-noise in each pixel (Appendix ~\ref{Sec.ErrorMapRatios}).}
\tablefoottext{b}{Averages and standard deviations of the mean derived from the ratios of the column densities in Table~\ref{table.columndensities}, measured from the extracted spectra at the positions indicated in Fig.~\ref{fig.maps}. Positions with upper limits are not considered. Similarly to the maps, parameters have been calculated unweighted, weighted with the standard deviation of each measurement, and weighted with the optical depth of the $^{13}$CO transition in Table~\ref{table.columndensities}.  See Sect.~\ref{sec.mapvsspectra} for a discussion on the differences observed between the maps and the spectra.
}}
\end{table*}

\subsection{Spectral analysis at selected positions}
\label{sec.spectra}

The positions studied in this work were selected by visually inspecting all the velocity channels in the \cdo~ data cube where individual flux density maxima were identified.
The coordinates of the positions identified are listed in Table~\ref{table.columndensities}. 
\cdo~ was selected for this task because of being bright while not affected by the effects of large optical depths, potentially affecting the brighter \tco, across the whole CMZ. The velocity structure and peak positions are similar for all isotopologues so the position selection is not biased towards  $^{18}$O enhanced regions.
This selection was not based on the previously identified GMC locations from existing high resolution studies of multi molecular studies of NGC~253 (see Appendix~\ref{Sec.positioncomparison}) since the purpose of this selection was to identify the maxima in our maps to enhance the detectability of the fainter isotopologues .

The integrated map of \tco~ in Fig.~\ref{fig.maps} shows the location of these positions.
Table~\ref{table.columndensities} also displays the projected galactocentric distance of each position referred to the bright compact radio source at $\alpha_{J2000}=00^h47^m33\fs17$, $\delta_{J2000}= -25^\circ17'17\farcs1$ \citep{Turner1985a,Ulvestad1997} assuming a distance of 3.5~Mpc to NGC~253. This distance is projected on the plane of the sky (i.e., not corrected for inclination) and may differ from the actual galactocentric distance, also among velocity components within a given position.
Some pairs of positions are very close in projected distance (i.e. positions 6 and 7) but they do correspond to the maximum flux densities of two distinct velocity components identified around those positions. 

\subsubsection{\tcdo~ line contamination}
\label{Sect.LineContamination}

\begin{figure}
\centering
\includegraphics[width=\hsize]{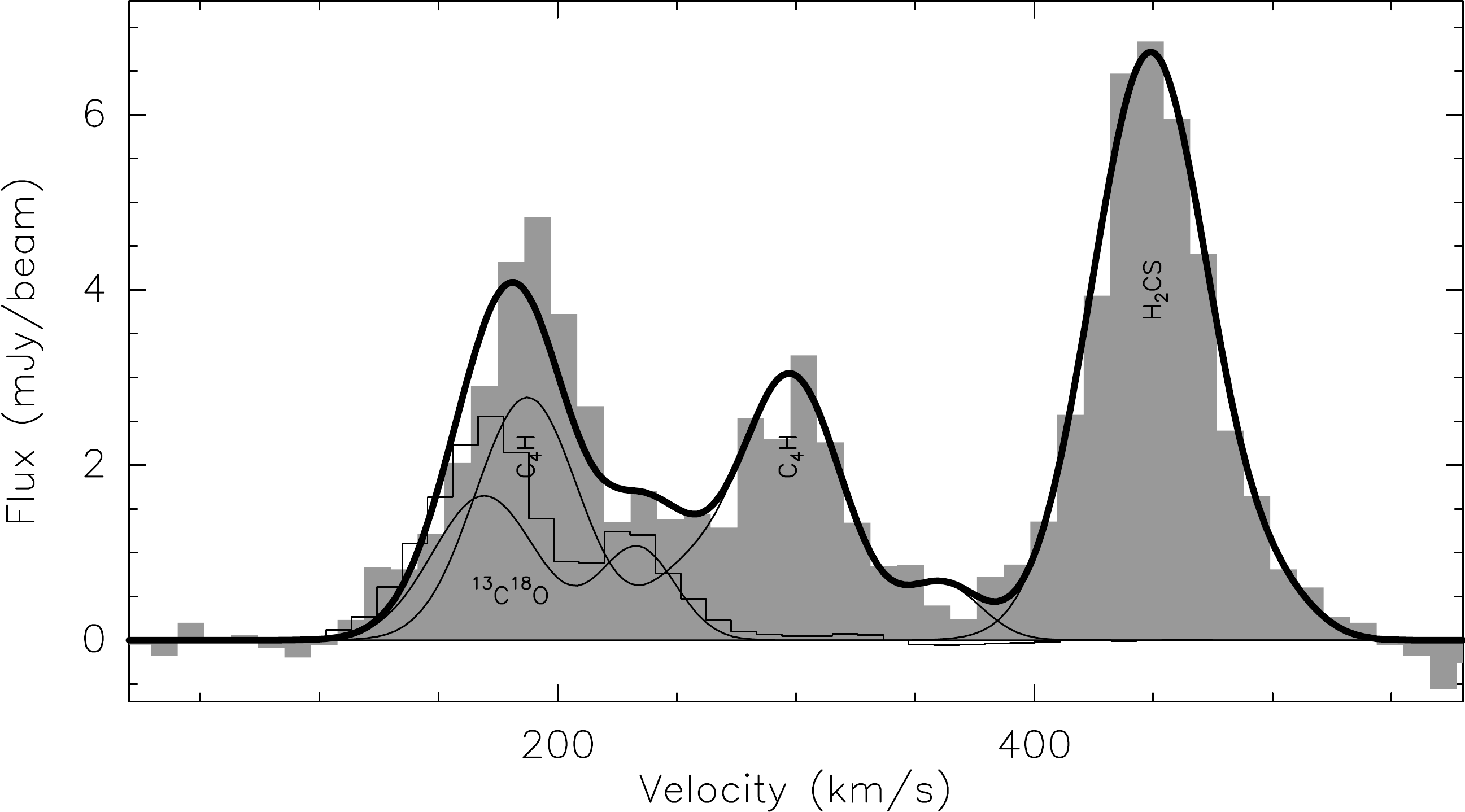}
   \caption{Spectra extracted at position 6 in Table~\ref{table.columndensities}. The grey filled histogram shows the spectrum around the \tcdo~ transition and the velocity scale refers to its rest frequency. The thick curve shows the overall fitted profile to the observations, while the thinner curves show the profiles of \tcdo, as well as the identified emission of the hyperfine structure of $\rm C_4H$ (see Sect.~\ref{Sect.LineContamination} for details) and the line of $\rm H_2CS$. 
The empty histogram shows the \tco~ profile (divided by 100) at the same position as a reference.
}
   \label{fig.linecontamination}
\end{figure}

\citet{Mart'in2010a} identified the contribution of the H$_2$CS $3_{1,2}-2_{1,1}$ transition at 104.616~GHz to the \tcdo~ line profile. The detection of H$_2$CS is actually confirmed by the detection within our observed bands of the $3_{0,3}-2_{0,2}$ line at 103.040~GHz with the expected relative flux density. 

However, in our high sensitivity data, we also identify two closer features to the \tcdo~($1-0$) transition, both emitting at lower frequencies (higher velocities). This is particularly obvious in the positions 6 and 7 in Table~\ref{table.columndensities} as shown in Fig.~\ref{fig.linecontamination} for position 6. As seen in that Figure, the observed profile is not similar to that of \tco, shown at a scaled down intensity for reference. The line shapes of \cdso~ and \cdo~, however, do match that of \tco. We note that these two positions are the most affected by this contamination.

In order to identify the origin of the emission blended to \tcdo~ we performed a preliminary fit to the spectra in these two positions using four Gaussian profiles (Note that the final fit in Fig.~\ref{fig.linecontamination} include more Gaussian profiles to account for the faint velocity components of the contaminants). The velocity of the first two were fixed to those of the two components observed in \tco, to account for the emission of \tcdo. The other two, accounting for the contaminant emission other than the already identified H$_2$CS line, had the velocity as free parameter but the width was fixed  to that of the brighter \tco~ component.
The features which are blended to \tcdo~ appear at $+23.0\pm1.3$~\kms, and $+129.2\pm1.4$~\kms~ with respect to the velocity of the brightest component of \tcdo, which sets the rest frequencies of the contaminating emission at $104703.3\pm0.5$~MHz, and $104666.3\pm0.5$~MHz, respectively. 

The spectral scans towards Sgr~B2 and the Orion hot cores by \citet{Turner1989} do indeed show an unidentified feature at 104696~MHz but nothing at lower frequencies, while the scan by \citet{Belloche2013} towards Sgr~B2(N) shows transitions of $\rm CH_3OCH_3$ and c-$\rm C_2H_4O$ around the frequencies of interest, overlapped with a number of vibrational transitions of $\rm C_2H_3CN$. Towards Sgr~B2(M), on the other hand, \citet{Belloche2013} detect just faint $\rm CH_3OCH_3$ emission around 104.7~GHz. However, based on our LTE synthetic spectrum of $\rm CH_3OCH_3$, a number of other transitions of this species should have been detected within our observed bandwidth, but they are not, which implies that $\rm CH_3OCH_3$ does not contribute significantly to the observed profile.
By using the spectroscopic information in the JPL and CDMS catalogs \citep{Pickett1998,Muller2001,Muller2005} we identified that our estimated frequencies agree well with those of the hyperfine transitions of $\rm C_4H$ at 104705~MHz ($11_{11}-10_{10}$) and 104666~MHz ($11_{12}-10_{11}$) with upper level energies $E_u=30$~K. Just recently, $\rm C_4H$ was detected for the first time in NGC~253 by \citet{Mangum2018} who reported higher energy transitions ($E_u=126$~K) in higher resolution observations. These are the first extragalactic detections of $\rm C_4H$ in emission, only previously reported towards the absorption system PKS~1830-211 \citep{Muller2011}. 

Despite the strong blending between \tcdo~ and the $11_{11}-10_{10}$ transitions of $\rm C_4H$ (low velocity C$_4$H transition in Fig.~\ref{fig.linecontamination}), we were able to account for the contribution of this transition to the observed spectral feature thanks to the relatively cleaner $\rm C_4H$ $11_{12}-10_{11}$ transitions (higher velocity C$_4$H transition in Fig.~\ref{fig.linecontamination}).
So, when analyzing the spectra (Sect.~\ref{Sect.LTEanalysis}), we considered as many velocity components of $\rm C_4H$ as those fitted to \tcdo. To de-blend the emission of \tcdo~ from that of C$_4$H we proceeded as follows. We fitted simultaneously the velocity components of \tcdo~ and C$_4$H. For the latter we used only the $11_{12}-10_{11}$ transitions (farther from the \tcdo~transition and therefore cleaner) to fit to the observed $\rm C_4H$ emission. Based on the fit to the $11_{12}-10_{11}$ transitions, the contribution of the $11_{11}-10_{10}$ lines (strongly blended with \tcdo) was then estimated from the expected relative intensities calculated with the spectroscopic parameters of the $\rm C_4H$ transitions involved. \tcdo~ emission was then fitted after removing the contribution of $\rm C_4H$. In some cases where the fainter velocity components of \tcdo~ were partially blended to the $11_{12}-10_{11}$ $\rm C_4H$ transitions, this fitting procedure required to be iterated until the fit to all components converged.
We also note that the fainter velocity components are subject to stronger blending uncertainties as illustrated in Fig.~\ref{fig.linecontamination}, resulting in large errors in the fits of both $\rm C_4H$ and \tcdo.
However, the values in Table~\ref{table.columndensities} can be considered as clean \tcdo~ column densities compared to the \tcdo~ integrated intensity map in Fig.~\ref{fig.maps}, where the $\rm C_4H$ emission is not removed (Sect.~\ref{sec.mapvsspectra}).

\begin{sidewaystable*}
\caption{Fitted parameters to the observed spectra}             
\label{table.columndensities}      
\centering          
\begin{tabular}{c  c  c  c  c  c  c  c  c  c  c  c  c  c  c  c}
\hline\hline
ID  &  R.A.$_{J2000}$  &  dec.$_{J2000}$  &  $d_{GC}$  &  $\rm v_{lsr}$  &  $\rm \Delta v_{1/2}$  &  $\tau_{^{13}\rm CO}$  &  $S(\rm^{13}CO)$  &  $N(\rm^{13}CO)$  &  $S(\rm C^{18}O)$  &  $N(\rm C^{18}O)$  &  $S(\rm C^{17}O)$  &  $N(\rm C^{17}O)$  &  $S(\rm^{13}C^{18}O)$  &  $N(\rm^{13}C^{18}O)$  &  $N(\rm C_4H)$ \\
  &  $00^h47^m$  &  $-25^\circ17'$  &  (pc)  &  ($km~s^{-1}$)  &  ($km~s^{-1}$)  &    &  ($mJy$)  &  ($log_{10}~cm^{-2}$)  &  ($mJy$)  &  ($log_{10}~cm^{-2}$)  &  ($mJy$)  &  ($log_{10}~cm^{-2}$)  &  ($mJy$)  &  ($log_{10}~cm^{-2}$)  &  ($log_{10}~cm^{-2}$) \\
\hline
1  &  $34\fs10$  &  $08\farcs65$  &  -257  &  91.9 (2.0)  &  21.7 (4.8)  &  0.05   &  52.8  &  16.2 (15.5)  &  7.9  &  15.4 (14.5)  &  1.4  &  14.6 (14.3)  &  <0.42  &  <14.2  &  <13.0 \\
   &   	  &   	  &     &  143.6 (1.0)  &  24.0 (2.4)  &  0.11   &  108.0  &  16.6 (15.5)  &  16.4  &  15.7 (14.5)  &  5.2  &  15.2 (14.3)  &  <0.41  &  <14.2  &  <13.0 \\
2  &  $34\fs03$  &  $12\farcs65$  &  -211  &  214.6 (1.3)  &  53.8 (3.0)  &  0.18   &  167.4  &  17.1 (15.8)  &  22.6  &  16.2 (14.8)  &  7.3  &  15.7 (14.8)  &  0.57  &  14.7 (13.9)  &  <13.0 \\
3  &  $33\fs73$  &  $16\farcs65$  &  -129  &  204.0 (1.7)  &  41.1 (4.0)  &  0.06   &  55.2  &  16.5 (15.4)  &  8.0  &  15.7 (14.6)  &  2.9  &  15.2 (14.3)  &  <0.16  &  <14.0  &  <13.3 \\
   &   	  &   	  &     &  290.1 (2.2)  &  47.3 (5.1)  &  0.05   &  45.7  &  16.5 (15.5)  &  7.2  &  15.7 (14.6)  &  1.4  &  14.9 (14.3)  &  <0.14  &  <14.0  &  <13.3 \\
4  &  $33\fs59$  &  $13\farcs15$  &  -117  &  112.6 (1.9)  &  22.4 (5.3)  &  0.06   &  54.5  &  16.2 (15.7)  &  8.1  &  15.4 (14.8)  &  2.7  &  14.9 (14.4)  &  0.25  &  14.0 (13.9)  &  <12.5 \\
   &   	  &   	  &     &  171.9 (1.5)  &  77.2 (3.7)  &  0.18   &  163.5  &  17.3 (15.9)  &  28.3  &  16.5 (15.0)  &  6.9  &  15.8 (14.6)  &  1.07  &  15.1 (14.2)  &  14.2 (14.1) \\
5  &  $33\fs62$  &  $14\farcs65$  &  -111  &  138.0 (2.3)  &  58.4 (5.4)  &  0.09   &  83.3  &  16.9 (15.8)  &  15.1  &  16.1 (15.0)  &  3.9  &  15.5 (14.5)  &  0.39  &  14.6 (14.1)  &  14.0 (13.5) \\
   &   	  &   	  &     &  201.3 (0.9)  &  47.3 (2.1)  &  0.22   &  202.1  &  17.2 (15.8)  &  33.1  &  16.4 (14.9)  &  8.6  &  15.7 (14.4)  &  1.11  &  14.9 (14.1)  &  13.9 (13.8) \\
6  &  $33\fs40$  &  $16\farcs15$  &  -55  &  169.2 (0.7)  &  51.0 (1.8)  &  0.27   &  244.2  &  17.3 (15.8)  &  52.6  &  16.6 (15.3)  &  11.5  &  15.9 (14.5)  &  1.65  &  15.1 (14.4)  &  14.6 (14.2) \\
   &   	  &   	  &     &  233.5 (1.3)  &  37.2 (3.0)  &  0.12   &  119.2  &  16.8 (15.7)  &  16.5  &  15.9 (15.2)  &  4.2  &  15.3 (14.5)  &  1.06  &  14.8 (14.3)  &  13.8 (14.1) \\
7  &  $33\fs33$  &  $16\farcs15$  &  -40  &  171.5 (0.5)  &  48.4 (1.2)  &  0.35   &  298.6  &  17.4 (15.7)  &  66.1  &  16.7 (15.3)  &  14.1  &  16.0 (14.4)  &  2.31  &  15.3 (14.3)  &  14.7 (13.4) \\
   &   	  &   	  &     &  235.3 (1.6)  &  44.2 (3.8)  &  0.09   &  91.2  &  16.8 (15.7)  &  13.8  &  15.9 (15.3)  &  3.3  &  15.3 (14.4)  &  1.13  &  14.9 (14.3)  &  <13.5 \\
8  &  $32\fs92$  &  $20\farcs65$  &  83  &  265.8 (0.9)  &  63.9 (2.1)  &  0.29   &  256.7  &  17.4 (15.9)  &  49.4  &  16.7 (15.2)  &  11.6  &  16.0 (14.7)  &  2.01  &  15.3 (14.2)  &  14.7 (13.9) \\
   &   	  &   	  &     &  369.9 (1.5)  &  41.0 (3.5)  &  0.13   &  123.1  &  16.9 (15.7)  &  19.2  &  16.0 (15.1)  &  4.0  &  15.3 (14.6)  &  0.71  &  14.7 (14.1)  &  <12.8 \\
9  &  $32\fs78$  &  $22\farcs15$  &  124  &  286.3 (0.8)  &  42.9 (1.7)  &  0.52   &  414.0  &  17.5 (16.1)  &  97.2  &  16.8 (15.3)  &  19.8  &  16.0 (14.7)  &  3.73  &  15.4 (14.4)  &  14.6 (13.3) \\
   &   	  &   	  &     &  378.1 (4.1)  &  34.0 (9.5)  &  0.07   &  66.7  &  16.5 (15.9)  &  12.2  &  15.8 (15.2)  &  2.4  &  15.0 (14.6)  &  0.82  &  14.7 (14.3)  &  <12.9 \\
10  &  $32\fs63$  &  $22\farcs15$  &  150  &  253.0 (6.2)  &  54.3 (9.1)  &  0.05   &  45.4  &  16.5 (15.9)  &  6.5  &  15.7 (14.6)  &  0.8  &  14.8 (14.5)  &  0.50  &  14.7 (14.2)  &  13.3 (13.7) \\
   &   	  &   	  &     &  291.9 (0.6)  &  35.0 (0.9)  &  0.24   &  213.4  &  17.1 (16.0)  &  45.1  &  16.4 (14.6)  &  10.0  &  15.7 (14.4)  &  1.84  &  15.0 (14.2)  &  14.0 (13.1) \\
   &   	  &   	  &     &  387.0 (0.7)  &  34.1 (1.6)  &  0.06   &  59.2  &  16.5 (15.1)  &  9.5  &  15.7 (14.5)  &  2.9  &  15.1 (14.4)  &  0.27  &  14.2 (14.0)  &  <12.9 \\
11  &  $32\fs30$  &  $21\farcs15$  &  211  &  252.5 (4.3)  &  70.2 (11.2)  &  0.02   &  24.5  &  16.4 (15.5)  &  5.1  &  15.7 (14.9)  &  1.5  &  15.1 (14.6)  &  <0.13  &  <14.2  &  <13.0 \\
   &   	  &   	  &     &  328.2 (0.4)  &  30.8 (0.9)  &  0.20   &  188.0  &  17.0 (15.4)  &  28.9  &  16.1 (14.7)  &  7.2  &  15.5 (14.5)  &  1.06  &  14.7 (13.9)  &  <12.9 \\
12  &  $32\fs44$  &  $27\farcs65$  &  245  &  332.0 (1.0)  &  60.1 (2.5)  &  0.08   &  76.0  &  16.8 (15.4)  &  13.8  &  16.1 (14.7)  &  3.6  &  15.5 (14.7)  &  0.47  &  14.7 (14.0)  &  <13.0 \\
13  &  $32\fs19$  &  $25\farcs65$  &  268  &  277.6 (0.7)  &  36.2 (1.7)  &  0.22   &  200.9  &  17.1 (15.7)  &  29.8  &  16.2 (14.7)  &  6.5  &  15.5 (14.3)  &  1.25  &  14.9 (14.0)  &  13.2 (13.1) \\
   &   	  &   	  &     &  323.9 (1.8)  &  33.2 (4.0)  &  0.08   &  78.8  &  16.6 (15.6)  &  14.7  &  15.8 (14.7)  &  3.5  &  15.2 (14.3)  &  0.49  &  14.4 (13.9)  &  <10.0 \\
   &   	  &   	  &     &  409.5 (2.7)  &  26.0 (6.4)  &  0.03   &  29.0  &  16.0 (15.4)  &  6.0  &  15.3 (14.6)  &  2.4  &  14.9 (14.2)  &  0.22  &  14.0 (13.8)  &  <10.0 \\
14  &  $32\fs04$  &  $28\farcs15$  &  320  &  299.9 (0.5)  &  50.1 (1.1)  &  0.21   &  194.0  &  17.2 (15.5)  &  30.2  &  16.3 (14.7)  &  9.8  &  15.8 (14.7)  &  0.90  &  14.9 (13.9)  &  13.5 (13.1) \\
   &   	  &   	  &     &  404.1 (4.9)  &  38.4 (11.5)  &  0.02   &  16.4  &  15.9 (15.4)  &  3.6  &  15.3 (14.7)  &  0.7  &  14.5 (14.7)  &  0.30  &  14.3 (13.9)  &  <12.5 \\
\hline
\end{tabular}
\tablefoot{$\rm v_{lsr}$  and  $\rm \Delta v_{1/2}$ refer to the velocity centroids and FWHM of the Gaussian components fitted for \tco. The fit assumes a $T_{ex}=15$~K and source size of $3''$ (see Sect.~\ref{sec.spectra} for details). Despite not being a fitted parameter (see Sect.~\ref{Sect.LTEanalysis}), the peak flux density of the $J=1-0$ transition of all isotopologues used in this analysis is also tabulated.
}
\end{sidewaystable*}

\subsubsection{\tcdo~ LTE analysis}
\label{Sect.LTEanalysis}
Spectra were extracted at each of the selected positions and modeled under the local thermodynamic equilibrium (LTE) assumption using MADCUBA\footnote{http://cab.inta-csic.es/madcuba/Portada.html} (Mart\'in et al in prep.).
With MADCUBA we fit a synthetic spectrum, calculated from input physical parameters and using the molecular spectroscopic parameters in the JPL and CDMS catalogs, to the observed line profiles. The free input parameters that can be fitted are the column density, excitation temperature, velocity, line width, and source size of each molecular species. Fig.~\ref{fig.linecontamination} shows a sample of the synthetic combined profile overlaid on top of the observed spectrum in one of the positions in this study. The results of the multi-Gaussian-component modeling to all selected positions are shown in Table~\ref{table.columndensities} where the obtained column densities derived for each isotopologue, as well as for $\rm C_4H$, are displayed. The peak flux densities of the observed CO isotopologues spectral lines (despite not being a direct fitted parameter, since MADCUBA does fit the column density) are also displayed in Table~\ref{table.columndensities} for reference.

Since this study is based on a single transition of each CO isotopologue, we need to make a number of assumptions: $i)$ all isotopologues have the same distribution, including velocity profile, $ii)$ they have a common excitation temperature of $T_{ex}=15~$K, and $iii)$ the source size is $3''$, matching the map resolution. Here we explain the implications of such assumptions.

The velocities and widths were fitted to the \tco~ profiles, and then set as fixed parameters when fitting the other isotopologues and $\rm C_4H$. This constraint was imposed to ensure consistency in the line profiles fitted, but in any case, letting these parameters free resulted in consistent values within the errors in the cases where the fit was possible both in terms of having enough signal-to-noise and not being significantly affected by line blending. Thus, fixing the velocity and width was only critical in the spectra with low signal-to-noise ratios and those of \tcdo~ where line blending was observed (Sect.~\ref{Sect.LineContamination}).

The excitation temperature was fixed to 15~K based on the rotational temperatures derived from multi-transition studies based on previous spectral scans \citep{Mart'in2006} as well on preliminary results from the ALMA multi-band spectral scan on NGC~253 (Mart\'in et al in prep.). Differences in the assumed excitation temperature have a minor impact on the derived column densities. For temperatures of 10 and 20~K, we estimate column densities $\sim5\%$ and $\sim10\%$ higher, respectively. However this would affect all isotopologues similarly if we assume they share similar excitation conditions, and therefore column density ratios would remain virtually unchanged. For temperatures below 10~K it is not possible to fit the intensity of observed profiles for any combination of column density and source size.

The column density and optical depth of the \tco~ transition are directly linked to the assumed source size, so these values would both increase under the assumption of a smaller source size. High resolution imaging by \citet{Ando2017} resulted in resolved GMCs of $\sim9$~pc ($0.5'' $) in size. However, in the case of position 9 in Table~\ref{table.columndensities}, the one with the largest measured optical depth (Table~\ref{table.columndensities}), we are not able to reproduce the observed flux density for source sizes $<2''$ unless significantly increasing the excitation temperature.
However as indicated above, multi-transition studies do not point towards such high excitation temperatures.
Thus we 
assume that the observed profiles stem from $2-3''$ averaged regions.
For a source size of $2"$, derived \tco~ column densities in Table~\ref{table.columndensities} will be underestimated by up to a factor of 3 in the brightest position 9. However, for all other positions, less affected by optical thickness, \tco~ column densities would be underestimated by a factor of $\lesssim 2.3$. 
On the other hand, the column density ratios in this study, would be underestimated by up to $40\%$ in position 9 for the ratios involving \tco, but we enter into the optically thick regime where fitted parameters strongly depend on the assumptions of source size and excitation temperature. For the optically thinner lines of sights and isotopologues, the derived ratios will be mostly unaltered.

Based on the column densities in Table~\ref{table.columndensities}, in Table~\ref{tab.isotopiratios} we show the average of the column density ratios and standard deviations calculated with the values obtained from all selected positions. These ratios do not include the positions where only upper limits to the \tcdo~ were obtained.
The averages and standard deviations are calculated: unweighted as a raw value; weighted by the standard deviation of individual measurements to get the average of the ratios measured at higher signal to noise or less blending; weighted by the optical depth of \tco~ as representative of the most massive clouds sampled.

\section{Discussion}
\subsection{Map vs spectra ratios: Optical depth and contamination effects.}
\label{sec.mapvsspectra}

From Table~\ref{tab.isotopiratios} we see how the average isotopologue ratios, calculated as the column density ratios derived from the spectral analysis (Sect.~\ref{Sect.LTEanalysis}) vary little and within the uncertainties when different weighting schemes are used. 
Although still consistent within the error bars (when unweighted averages are considered), these values from spectral analysis are higher than those measured from the averaging of the maps. These differences can be considered significant since they are both measured with the same dataset.
Here we do discuss why these difference can be attributed to a combined effect of both optical depth and line contamination affecting the lines in the measured ratio.

As explained above, \tco~ is affected by significant optical depth (Sect.~\ref{Sect.LTEanalysis}) while \tcdo~ is contaminated by $\rm C_4H$. Thus, we observe that the optical depth affected ratio \tco/\cdo~ is $38\pm6\%$ smaller in the averaged map than from the spectra (considering unweighted values from Table~\ref{tab.isotopiratios}). On the other hand, the contamination affected ratio \cdo/\tcdo~ is observed to be $22\pm9\%$ smaller. Subsequently, the \tco/\tcdo~ ratio, affected by both effects, should differ by approximately the multiplication of both factors, and therefore be $70\pm40\%$ smaller and, indeed the observed difference between the spectra and map derived average is $90\pm40\%$.

This explanation is further supported by the fact that the derived averages diverge even further from those derived with the spectra when weighted by the standard deviation or the signal-to-noise ratio per pixel. These weightings do favour the brighter regions, which are those that will be either more opaque and/or more contaminated by $\rm C_4H$.

On the other hand, the \cdo/\cdso~ ratio is expected not to be affected by line saturation and we observe that is the only ratio which is higher in the averaged maps compared to the spectra and which show little variation with the weighted averages from the maps. Therefore, the differences between the values derived from the maps and the spectra can be actually attributed to actual spatial variations. We note that the small variation derived from the weighted average in the \cdo/\cdso~ratio, could still be due to significant optical depth in the \cdo~ emission.

Based on the arguments above we can safely consider that the values derived from the spectra are the most accurate optical depth and contamination corrected values as proxies of the elemental isotopic ratios in NGC~253.
While we will adopt the $\sigma$-weighted ratios from the spectra for the discussion on stellar processing in Sect.~\ref{sec.processedgas}, we will use the unweighted values as reference for comparison with previous single-dish data in the literature in Sect.~\ref{Sec.IsotopicRatios}.

\subsection{Spatially resolved carbon and oxygen ratios with optically thin tracers}
\label{Sec.IsotopicRatios}

Based on the observations presented in this paper, we can for the first time derive spatially resolved  carbon and oxygen isotopic ratios based on the rarer carbon monoxide isotopologues. This is under the assumption that the column density ratio of the molecular isotopologues does reflect the actual atomic isotopic ratio.

As seen in Figs.~\ref{fig.mapsratios} and ~\ref{fig.isotopicplots}, the measured isotopic ratios vary significantly across the central molecular zone in NGC~253. Based on the column density values determined from the spectra (Table~\ref{table.columndensities}), we observe ratios ranging from 10 to 34 for \dtc, 45 to 270 for \dsdo, and to a lesser extent 2.7 to 8.5 for \dods.
Still, on average our high resolution results are about a factor of two below those typically assumed in the nuclei of galaxies \citep{Wilson1994,Henkel1994a,Wang2004}.

As indicated in the introduction the selection of optically thin tracers does minimize the effect of selective photodissociation. However, if \tcdo~ should be affected by a higher photodissociation rate \citep{Visser2009}, the ratios derived in this paper would be even lower. However, we have no evidences to support such effect globally within the central region of NGC~253.

Here we put our results in the context of previously determined isotopic ratios towards NGC~253.

\subsubsection{\dtc}
The single dish ratio \cdo/\tcdo$\gtrsim60$ reported by \citet{Mart'in2010a} differs by a factor of 3 from the ratio reported in this work.
We note that the \cdo~ peak temperature reported by \citet{Mart'in2010a} is $75\%$ brighter than that reported by \citet{Aladro2015}. Additionally, \tcdo~ and \cdo~ where observed at different positions, SD~B and SD~A, respectively. As explained in Sect.~\ref{Sec.SpatialFiltering}, \tcdo~ could be $25\%$ brighter towards the SD~A position. Thus, if these uncertainties do align in the same direction, the ratio derived by \citet{Mart'in2010a} could result in a value as low as $\gtrsim35$, closer to the value derived with other tracers \citep{Henkel2014}, but still higher than the value derived in this paper.
However, the single dish value from \citet{Mart'in2010a} is uncertain because of the unknown systematic errors resulting from the use of heterogeneous datasets.

The value derived from our optically thin observations is a factor of 2 lower than the ratio typically assumed for galaxies \citep{Wilson1994,Henkel2014} and a factor of 2-4 below the limits based on CCH \citep{Mart'in2010a}, in this case derived from the homogeneous spectral scan dataset by \citet{Mart'in2006}.

As we discuss in Sect.~\ref{sec.processedgas}, this difference may be real and a result of spatially distinct molecular gas component with different degrees of stellar processing.

\subsubsection{\dsdo}
The value quoted in the literature \citep{Wilson1994} of $\sim200$ is derived from the \tco/\cdo~ ratio reported by \citet{Sage1991} in a sample of three galaxies (NGC~253, IC~342, M~82), and multiplied by the assumed \dtc$\sim40$ \citep{Henkel1993}.
For NGC~253, \citet{Sage1991} reported \tco/\cdo$=4.9\pm0.5$,  which is good agreement with the value of $4.5\pm1.7$ derived from our map measurements in Table~\ref{table.columndensities}.
If we do use our derived \dtc~ to multiply the measurement by \citet{Sage1991}, we would obtain a ratio \dsdo$=110\pm30$, which is consistent with our optical depth/contamination corrected (Sect.~\ref{sec.spectra}) measurement of \tco/\tcdo.
Thus, the discrepancy between the \dsdo~ ratio reported in this work and that of \citet{Wilson1994} does reside on the large scale \dtc~ ratio differing from that measured at high resolution.

\subsubsection{\dods}
\citet{Sage1991} reported a \dods~ of $10\pm2.5$ based on \cdo/\cdso~$J=2-1$ observations towards NGC~253. The discrepancy here can be attributed to the low signal-to-noise ratio of their single dish data. The higher sensitivity observations by \citet{Aladro2015} of the $J=1-0$ yield a ratio of $7.1\pm0.4$, also in good agreement with the map derived value of $6.5\pm3.7$ in Table~\ref{tab.isotopiratios}.

\subsection{Isotopic ratio variations across the CMZ of NGC~253}

\begin{figure*}
\centering
\includegraphics[width=\hsize]{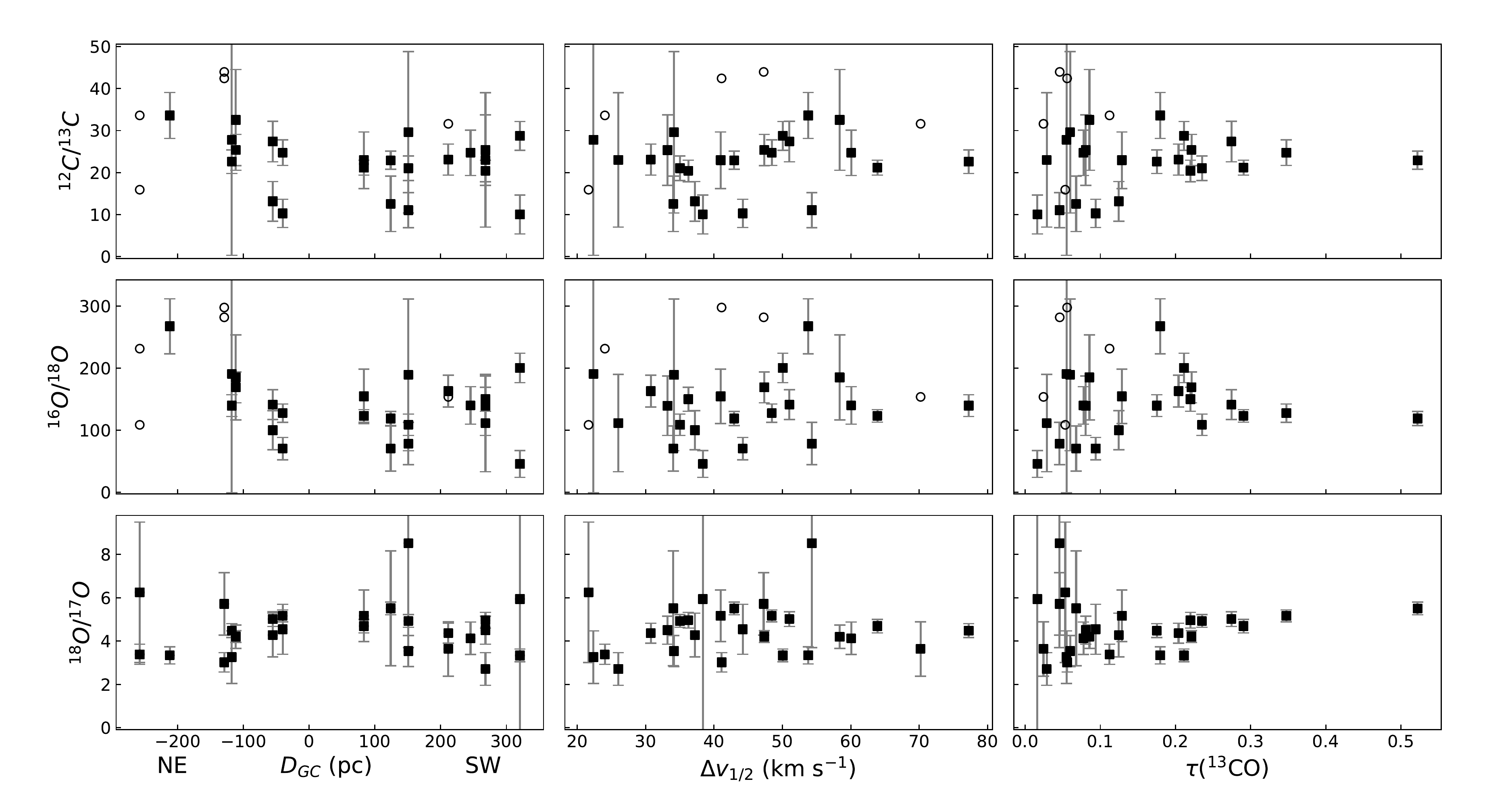}
   \caption{Measured isotopic ratios as a function of galactocentric distance, full width at half maximum of the fitted profile and optical depth of the \tco~transition. Ratios are calculated based on the column density ratios in Table~\ref{table.columndensities}. Lower limits ($3~\sigma$) to the ratios due to non-detection of \tcdo~ are displayed as open circles.}
   \label{fig.isotopicplots}
\end{figure*}

Since the ALMA observations allowed us to spatially resolve the central molecular zone of NGC~253 we explore the variations of isotopic ratios among the individual identified sources across this region.

Within our Galaxy, a clear gradient as a function of the galactocentric distance is observed in the carbon, oxygen and nitrogen isotopic ratios \citep{Gardner1979,Henkel1982,Langer1990,Wilson1994,Goto2003,Milam2005,Wouterloot2008,Zhang2015}. The left panels in Fig.~\ref{fig.isotopicplots} show the variation of estimated isotopic ratios as a function of the linear projected distance measured from the assumed galactic center position (see Sect.~\ref{sec.spectra}).
We do not observe an obvious gradient in our data as a function of galactocentric distance. However, we note that, within the Galaxy, this gradient is observed at distances ranging 3 to 10~kpc, while our observations in NGC~253 cover the inner $\sim300$~pc. In fact, considering the dispersion of the observations at a given Galactocentric distance, such gradient, should it continue towards the very central region, might not be distinguished within the central few hundred parsecs of the Galaxy as is the case in NGC~253. Most of these studies present only one observational point towards the Galactic center ranging \dtc~$=17\pm7$ \citep[where the error includes the variation of the ratio measured with different molecular tracers used,][]{Langer1990,Milam2005} with a standard value generally adopted of $\sim20$ \citep{Wilson1994}. Moreover, the study by \citet{Gardner1982} shows that the low ratio of ratio \dtc~$\sim15\pm4$, as measured with H$_2$CO, is relatively homogeneously observed across the central molecular zone and not exclusively from individual sources like Sgr~A or Sgr~B2. This is similar to the relatively homogeneous ratios observed within the central molecular zone of NGC~253. Similarly, oxygen ratios show no variations with distance from the center of NGC253 within a radius of ~300 pc.
Similarly, we do not observe obvious gradients in \dsdo~ and \dods~ ratios, with observed ratios within $25\%$ of the average (see Table~\ref{tab.isotopiratios}).

In order to further investigate the relative variations of the isotopic ratios among molecular components, we plot the measured ratios as a function of the observed line width. Line width was taken as a proxy of the virial mass of the unresolved clouds, because of the lack of an a priory knowledge of their individual sizes. Then central panels in Fig.~\ref{fig.isotopicplots} show no apparent trend with the line width on any of the isotopic ratios.

The right panels in Fig.~\ref{fig.isotopicplots} show the ratio dependency on the measured optical depth of \tco~ ($^{13}\tau$) at each position. This optical depth assumes the same source size for all sources. The implications of this assumption have already been discussed in Section~\ref{sec.spectra} and translate into an uncertainty in the x-axis of this dependency, where points might be displaced by up to a factor of $\sim2$ towards higher optical depths for some positions. 
Once again, no obvious trend is observed. However the most uncertain ratios are found towards the weaker molecular components (those with lower optical depths and therefore column densities) resulting in larger error bars and uncertain lower limits in the \dtc~ and \dsdo, due to the strong blending with $\rm C_4H$ (Sect.~\ref{Sect.LineContamination}. This is also reflected in larger uncertainties due to the limited signal-to-noise ratios in \dods, unaffected by blending.

\begin{figure}
\centering
\includegraphics[width=\hsize]{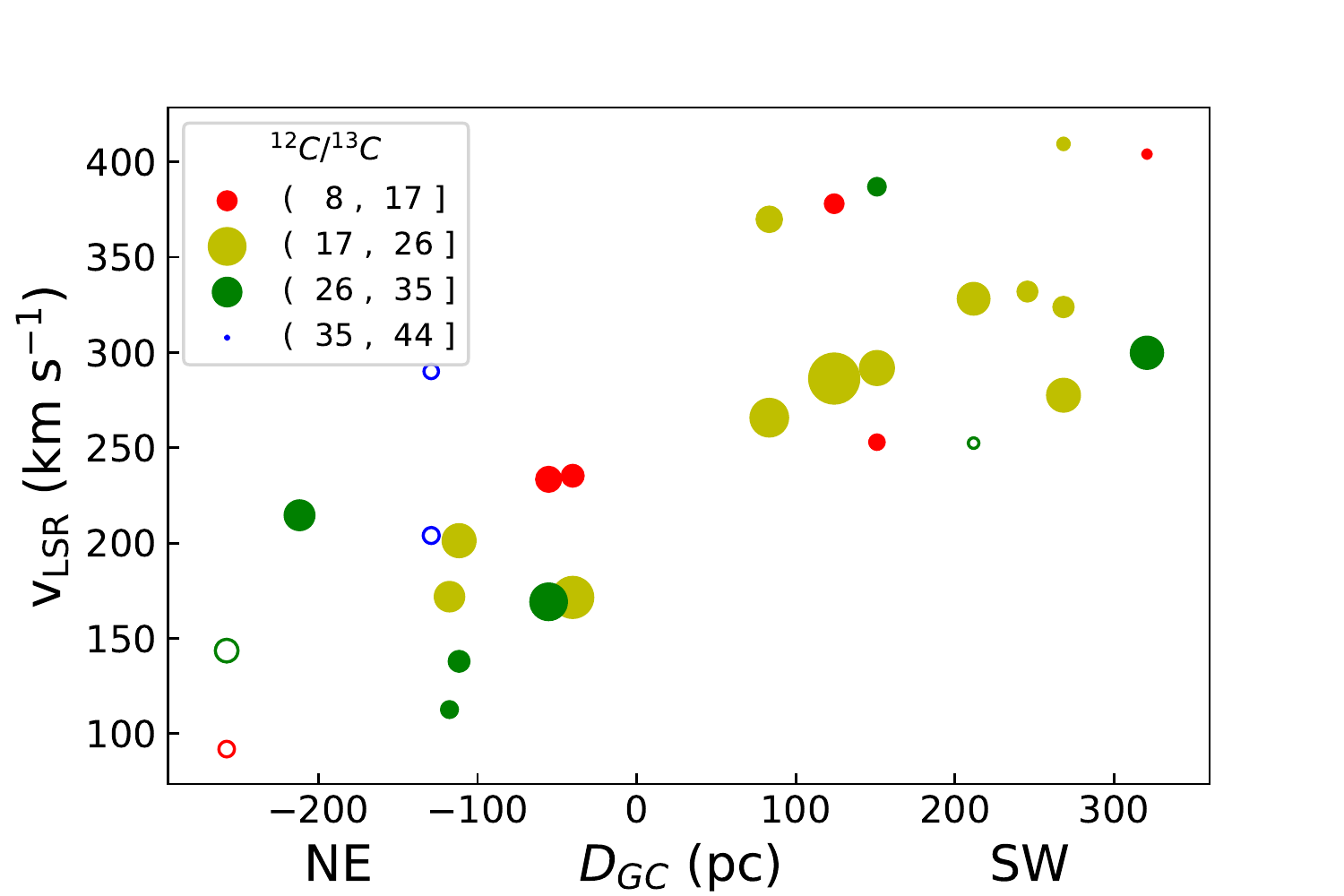}
\includegraphics[width=\hsize]{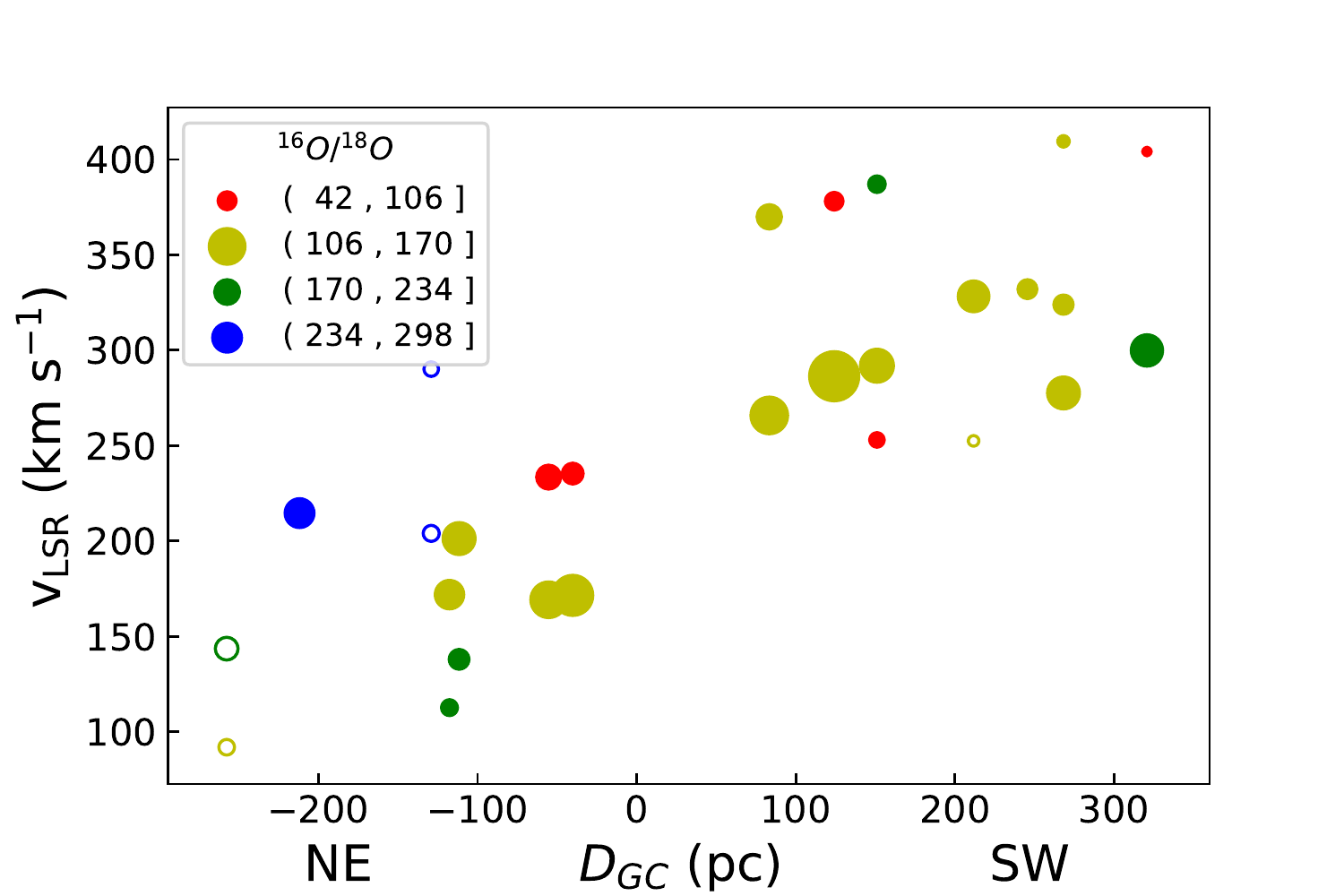}
   \caption{Color coded carbon (top) and oxygen (bottom) isotopic ratios as a function of the distance to the galaxy center and separated by the measured velocity of each fitted component. Color coding ranges were selected to distinguish four quartiles between the minimum and maximum values measured in all components. The size of the points is related to the optical depth of the \tco. Open symbols represent lower limits.}
   \label{fig.isotopicplotskinematics}
\end{figure}

Additionally, in Fig.~\ref{fig.isotopicplotskinematics} we show the variations of the carbon and oxygen isotopic ratios across all the positions selected, where the different velocity components are separated in four colour groups, according to the measured ratios, where also the optical depth of \tco is coded with the size of the points. Higher granularity in the color coding does not provide a more accurate picture given the errors in the ratios.

Different from the situation in our Galactic center \citep[Fig.~2 in ][]{Gardner1982}, here we can assume that all velocity components are located within the central molecular zone. However our spatial resolution is not high enough to provide an accurate picture of their locations. Higher resolution observations will provide a more accurate picture of the isotopic distribution.

We observe an overall homogeneity of the measured isotopic ratios where most of the gas is around the averaged reported values. Only a few low column density velocity components show ratios significantly above or below the average (blue and red, respectively, in Fig.~\ref{fig.isotopicplotskinematics}).

\subsection{Stellar processed gas in the galactic center of NGC~253}
\label{sec.processedgas}

The relatively large \dtc~ratio observed in NGC~253 (and other starbursts galaxies) as compared to that in the Galactic center has  been claimed as the result of infalling unprocessed material into the central region. Such accretion has also been claimed towards the central molecular zone of the Milky Way through the study of isotopic enrichment \citep{Riquelme2010}.
Our results show that at high resolution the degree of stellar processing of the molecular gas in the central molecular zone of NGC~253, as measured by carbon monoxide isotopologues, appears to be similar to the one in the Galaxy.

Our \dtc~ ratio of $21\pm6$ is well in agreement with the value of $24\pm1$ in the Galactic center, measured with the sames isotopologue transitions of CO \citep{Langer1990}, while the \dsdo~ ratio of $130\pm40$ is about half of the 250 measured there \citep{Wilson1994}. However, the value in NGC~253 would result in a good agreement if we extrapolate the galactocentric trend observed across the large scale disk of  the Galaxy \citep[Fig.~2 in ][]{Wilson1994}.
Similarly, the \dods~ ratio of $4.5\pm0.8$ is about $50\%$ larger than the measurements by \citet{Zhang2015} 
towards our Galactic center. 
The smaller \dsdo~ and larger \dods~ observed ratios may point out towards an actual enhancement of $^{18}$O as compared to the central molecular zone of the Milky Way.
This might be attributed to the fast enhancement of $^{18}$O by massive stars, while $^{13}$C and $^{17}$O are more slowly injected by low- and intermediate-mass stars \citep{Zhang2018}.
Indeed, \citet{Zhang2015} find a difference in the \dods~ ratio between the Galactic center and the molecular clouds in the disk, which implies differences in the gas phase injection of these isotopes.

Our result at high resolution implies that the low resolution ratios measured with single dish observations do actually reflect an average global value that may include two well separated molecular components, one significantly processed by the past star formation in the nuclear region and a less processed component likely infalling from the outer disk that will presumably be feeding the future star formation in the region.

In this scenario, the recent accurate optical depth corrected \dtc~ measurement of $\sim40$ based on CN observations by \citet{Henkel2014} would actually be the average of the processed material in the central region (\dtc$\sim20$) plus a molecular component with a higher \dtc~ ratio. 
If we consider the recovered fluxed estimated in Sect.~\ref{Sec.SpatialFiltering}, the filtered out molecular component would correspond to $\sim70\%$ of the \tco~ single dish emission, as well as $40-55\%$ of that of \tcdo. This yields isotopic ratios for the filtered component of \dtc$\sim50-70$, in order to explain the global averaged single dish carbon isotopic ratio.
This extended molecular component might be actually the one traced by CCH \citep{Mart'in2010a}, where ratios $>56$ and $>81$ were estimated based on the non-detection of $^{13}$CCH, and the stacked spectra of $^{13}$CCH+C$^{13}$CH, respectively.
In fact, the observation of CN obtained as part of the ALCHEMI line survey towards NGC~253 (Mart\'in et al. in prep.) show an optical depth well in excess of the $\tau\sim0.2-0.5$ obtained by single dish data \citep{Henkel2014}, which would further support the fact that single dish data are indeed the result of the averaging of an optically thin extended molecular component plus a compact optically thick one.

A similar calculation cannot be estimated for the extended \dsdo~ ratio since the single dish global value is actually derived through assumptions on the \dtc~ratio as explained in Sect.~\ref{Sec.IsotopicRatios}.

While large scale extragalactic unresolved observations might still make use of commonly assumed single dish derived ratios, high resolution observations may need to consider the multicomponent nature of the molecular clouds in the central region of galaxies. 
As an example, we note that the HCN/H$^{13}$CN and HCO$^+$/H$^{13}$CO$^+$ line ratios ranging $10-15$ obtained at $2''$ resolution \citep{Meier2015} are actually close to the \dtc~ isotopic ration derived in this paper. Thus their derived optical depths of $\tau\sim5-8$ for HCN and HCO$^+$ would actually be significantly overestimated. If we recalculate the optical depths for these two dense gas traces assuming our derived \dtc$\sim21$, it results in moderate optical depths of $\tau\sim0.7-1.7$.

Our results also shed some doubts on the feasibility of inferring the history of stellar nucleosynthesis and the characteristics of the IMF in galaxies based on  global isotopic ratios \citep{Romano2017}. Actually the molecular gas affected by the stellar nucleosynthesis in the central region might be masked or diluted by the infalling of gas from less processed regions, as it appears to be the case in NGC~253, or as result of galactic interactions.
\citet{Zhang2018} recently reported a potential trend of the \tco/\cdo~ line ratio as a function of the infrared luminosity in a sample of extragalactic sources. This trend is claimed to be linked to the stellar initial mass function (IMF). In that scenario, the lowest ratios of $\sim1$ would be the result of a top-heavy stellar IMF.
However, once again this is based on a single ratio at low spatial resolution, while our study shows the significant difference between the direct measurements from the maps and the optical depth corrected values towards the selected positions (Sect.~\ref{sec.spectra}).

\section{Conclusions}
The main results from the observations of the rarer isotopologues of carbon monoxide are:
\begin{itemize}
\item{For the first time we present unambiguous and spatially resolved detections of the double substitution of carbon monoxide \tcdo~ in the extragalactic ISM of the starburst NGC~253.}
\item{Carbon and oxygen isotopologue ratios have been derived with optically thin tracers at high spatial resolution resulting in lower values than previously obtained with single dish low resolution data. Our derived ratios take into account and correct the contamination of \tcdo~ by $\rm C_4H$ and the moderate optical thickness of \tco.}
\item{The deduced \dtc$\sim21$ agrees with the value measured towards the center of the Milky Way, which can be understood in terms of a similar degree of nuclear processing from stellar nucleosynthesis. Both \dsdo$\sim130$ and \dods$\sim4.5$ are well below and above those measured in the Galactic center, respectively, which points out to a $^{18}$O enhancement by massive stars and a slower injection of $^{13}$C by low- and intermediate-mass stars.}
\item{Differences from the ratios observed with single dish telescopes appear to present evidence of a multicomponent scenario with molecular gas highly processed in the central region of NGC~253 and unprocessed gas claimed to be infalling from the outer regions of the galaxy.}
\item{No obvious gradients are found as a function of the distance to the center out to galactocentric radii of $\sim300$~pc. 
}
\end{itemize}
 
\begin{acknowledgements} 
SM want to thank the valuable discussion with V. Rivilla, J. Mart\'in-Pintado, and Laura Colzi on the results presented in this paper.
This paper makes use of the following ALMA data: ADS/JAO.ALMA\#2016.1.00292.S. ALMA is a partnership of ESO (representing its member states), NSF (USA) and NINS (Japan), together with NRC (Canada), MOST and ASIAA (Taiwan), and KASI (Republic of Korea), in cooperation with the Republic of Chile. The Joint ALMA Observatory is operated by ESO, AUI/NRAO and NAOJ.
\end{acknowledgements} 

\bibliographystyle{aa}
\bibliography{13C18ONGC253}

\appendix
\section{Error and signal to noise in line ratio maps}
\label{Sec.ErrorMapRatios}

\begin{figure}
\centering
\includegraphics[width=\hsize]{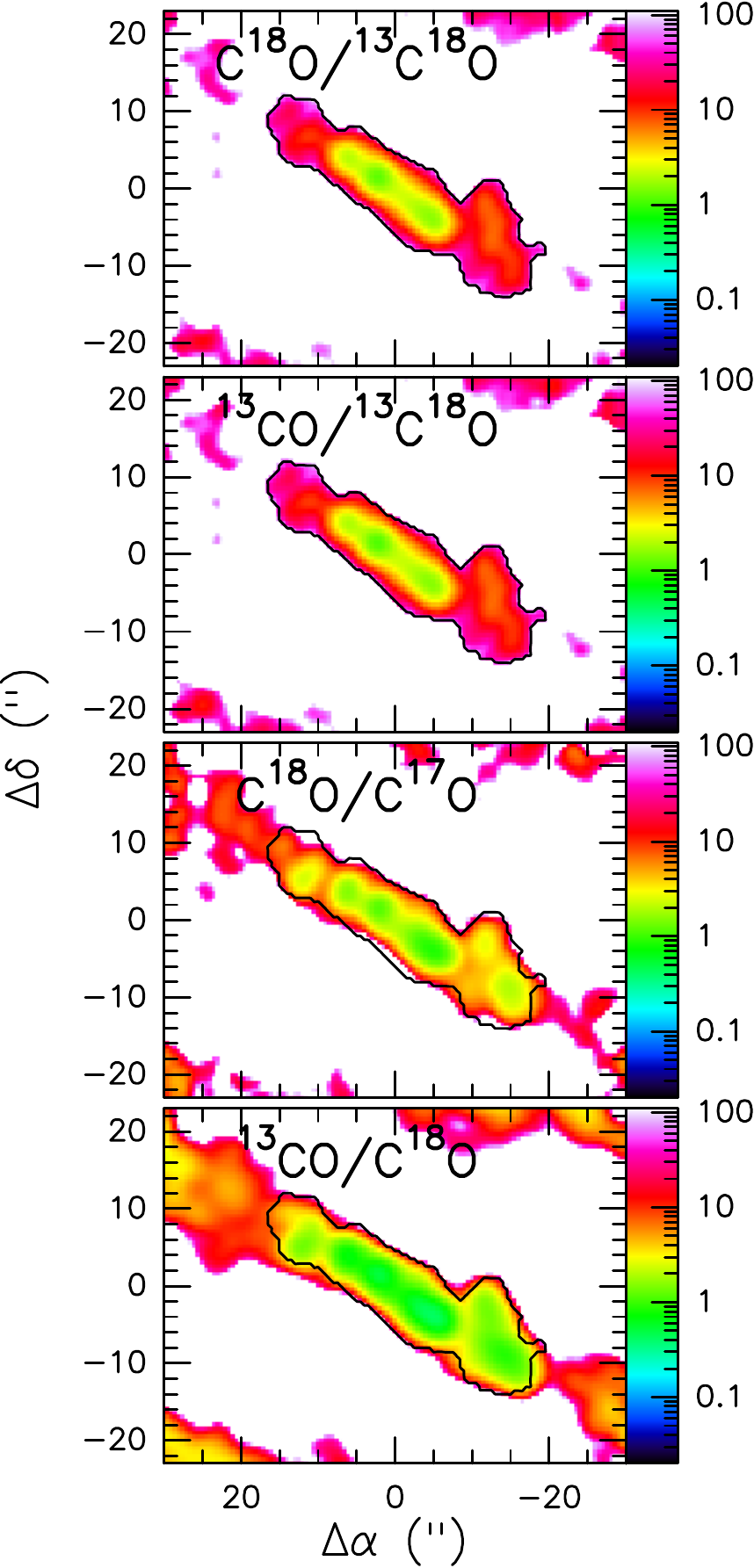}
   \caption{
   Accuracy of the ratio maps shown in Fig.~\ref{fig.mapsratios} in percentage as defined in Eq.~\ref{eq.accuracy}. The black contour is similar to the one in Fig.~\ref{fig.mapsratios}.
   }
   \label{fig.errormapsratios}
\end{figure}

   
While Fig.~\ref{fig.maps} shows the ratio $R=M_1/M_2$ between the corresponding maps $M_1$ and $M_2$, Fig.~\ref{fig.errormapsratios} displays the ratio between the propagated error of the ratio map over the ratio map in percentage as 
\begin{equation}
    \frac{\sigma_R}{R} (\%)=\frac{100}{R}~\sqrt{
    \left ( \frac{\sigma_{M_1}}{M_2}\right )^2 +
    \left ( \frac{\sigma_{M_2}~M_1}{M_2^2}\right )^2 
    }
    \label{eq.accuracy}
\end{equation}

where we observe how the central region, as expected, has errors of $\sim1\%$ rising to a few 10~\% towards the edge of the region considered (see Sect.~\ref{sec.ratiomaps}).

\begin{figure}
\centering
\includegraphics[width=\hsize]{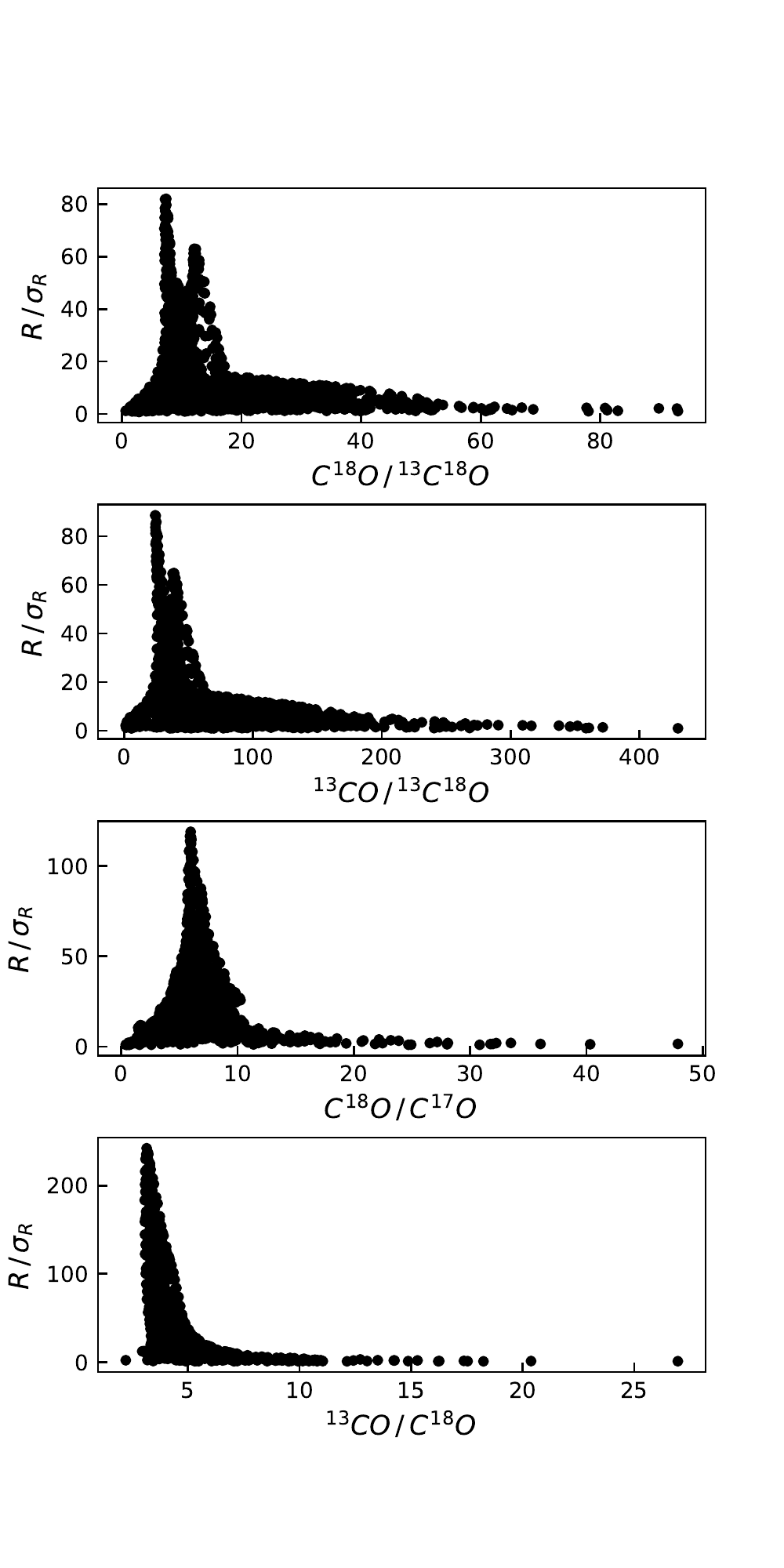}
   \caption{Signal-to-noise ratios per pixel in the ratio maps of Fig.~\ref{fig.mapsratios} as a function of the line ratio.}
   \label{fig.snrdistribution}

\end{figure}

Fig.~\ref{fig.snrdistribution}, on the other hand, shows the signal to noise ratio per pixel of the ratio maps ($\rm R/\sigma_{R}$) as a function of the ratio. The effect of line contamination by C$_4$H (Sect.~\ref{Sect.LineContamination}) in the \cdo/\tcdo~ and \tco/\tcdo~ is clearly seen a the double peak structure and a broad tail both at high signal to noise as a result of the different distributions between the CO isotopologues and C$_4$H. This is different from what is observed in the \cdo/\cdso~ and \tco/\cdo~ ratios which show a single peaked distribution. The effect of the significant \tco~ optical depth is evidenced by the \tco/\cdo~distribution being skewed towards lower values, different from what we observe in the optically thin ratio \cdo/\cdso~ more symmetrical distribution. Similarly the two high signal-to-noise components in \tco/\tcdo~ are also pushed together by the effect of optical depth. See Sect.~\ref{sec.mapvsspectra} for further discussion of both optical depth and line contamination effects on ratio maps based on the differences observed with the ratios from spectral analysis.

\section{Selected positions: Comparison to other works}
\begin{figure}
\centering
\includegraphics[width=\hsize]{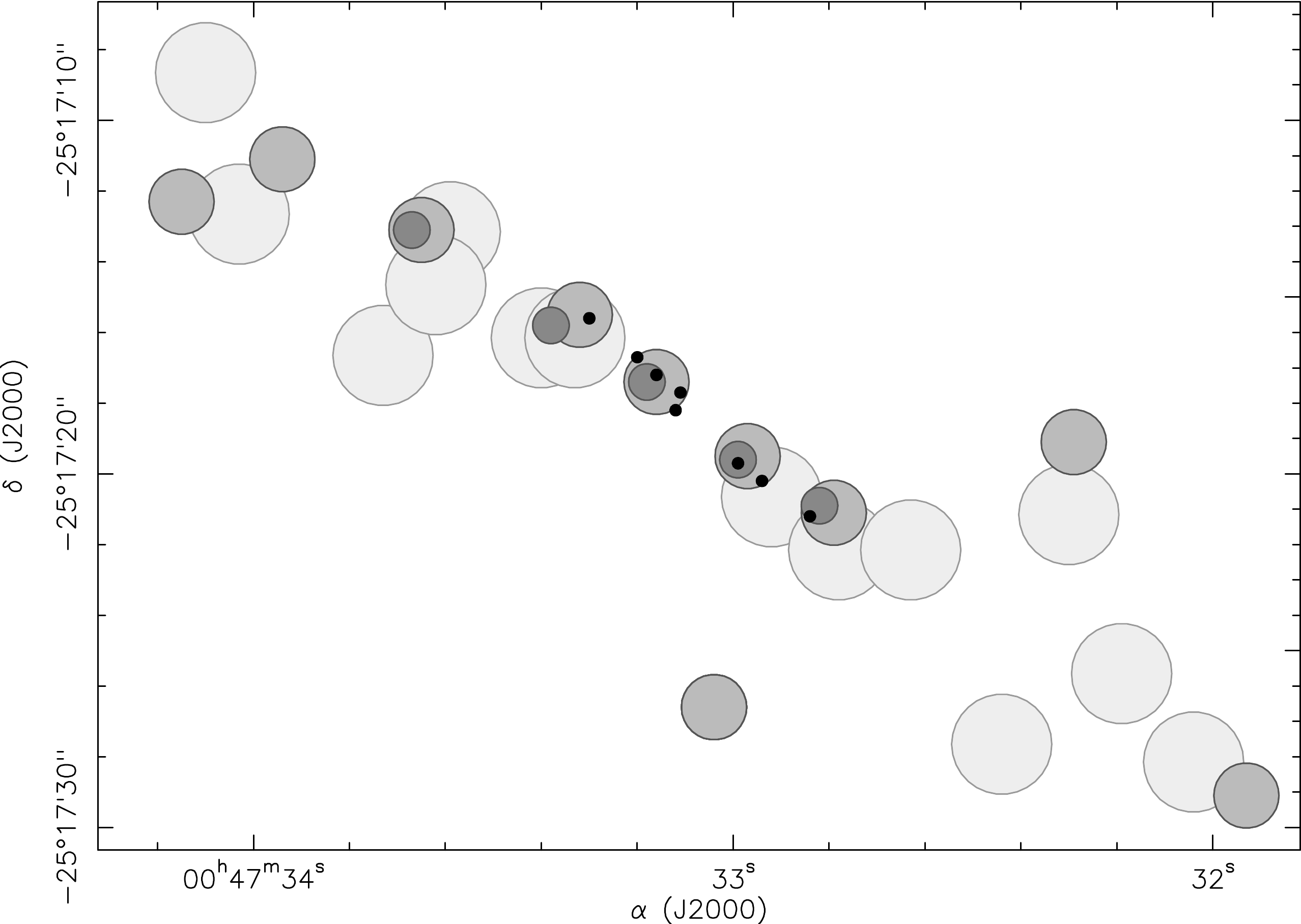}
   \caption{Location of the positions analyzed in this paper ($3''$, light grey circles), compared to the positions from \citet{Meier2015} ($2''$, grey circles), \citet{Sakamoto2011} ($1.1''$, dark grey circles), and \citet{Ando2017} ($\sim0.37''$, black dots). The size of the circles represent the resolution of the observations in these studies.
   }
   \label{fig.positions}

\end{figure}

Positions in this work have been based on observed maxima in the \cdo~ channel maps (see Sect.~\ref{sec.spectra}) and not from those in previous high resolution studies \citep{Sakamoto2011,Meier2015,Ando2017,Leroy2015,Walter2017}. Fig.~\ref{fig.positions} shows the positions selected in this paper compared to those by \citet{Meier2015} aiming at sampling different regions in NGC~253 CMZ, those of \citet{Sakamoto2011} identifying the brightest spots in the integrated intensity maps, and those of the high resolution study by \citet{Ando2017} identifying the peak continuum sources in the very inner region. The different angular resolutions of these studies in the literature are also shown in this Fig.~\ref{fig.positions}. 

\label{Sec.positioncomparison}

\end{document}